\documentclass[
 twocolumn,
showpacs,preprintnumbers,
showkeys,
 amsmath,amssymb,
 aps,
 prstab,
]{revtex4-1}

\usepackage{graphicx}
\usepackage{dcolumn}
\usepackage{bm}
\usepackage{comment}

\begin{document}


\title{High-efficiency broadband THz emission via diffraction-radiation cavity}

\author{Yosuke Honda}
 \email{yosuke@post.kek.jp}
\author{Miho Shimada}%
\author{Alexander Aryshev}%
\author{Ryukou Kato}%
\author{Tsukasa Miyajima}%
\author{Takashi Obina}%
\author{Ryota Takai}%
\author{Takashi Uchiyama}%
\author{Naoto Yamamoto}%
\affiliation{%
High Energy Accelerator Research Organization (KEK), 1-1 Oho, Tsukuba, Ibaraki, Japan
}%

\date{\today}

\begin{abstract}
Accelerator-based terahertz (THz) radiation has been expected
to realize a high-power broadband source.
Employing a low-emittance and short-bunch electron beam at a high repetition rate,
a scheme of coherent diffraction-radiation
in an optical cavity layout is proposed.
The scheme's stimulated radiation process between bunches
can greatly enhance the efficiency of the radiation emission.
We performed an experiment with a superconducting linac 
constructed as an energy recovery linac (ERL) test facility.
The electron beam passes through small holes in the cavity mirrors
without being destroyed.
A sharp THz resonance signal,
which indicates broadband stimulated radiation
correlated with beam deceleration,
was observed while scanning the round-trip length of the cavity.
This observation proves the efficient beam-to-radiation energy conversion
due to the stimulated radiation process.
%
%
\end{abstract}


\maketitle


\section{Introduction}
Light sources have played important roles in the progress of science in various fields.
The terahertz (THz) spectrum range, which is usually defined from 0.3 to 3 THz,
spans the technological gap between microwave and optical technologies \cite{tonouchi}.
THz light sources are the most immature over a wide spectrum of light.
As it corresponds to a photon energy of 1.2 to 12 meV 
and a temperature of 14 to 140 K,
the natural environment at room temperature is filled with THz-scale phenomena.
For example,
the excitation energy of molecular dynamics,
such as the rotation and oscillation of a molecule and lattice,
corresponds to the THz scale.
Development of THz sources is necessary for investigating such dynamics.
For the production of high-power radiation, 
the efficiency of THz radiation production is the key for the source development.

A table-top Fourier-transform infrared spectroscopy system (FT-IR) \cite{ftir} 
has been developed utilizing a thermal source.
The spectrum of the thermal source is determined by black-body radiation,
which peaks in the near-infrared range,
and its power in the THz range is quite small and unstable.
A time-domain spectroscopy (TDS) system has been developed
utilizing photoconductive antenna or nonlinear crystal.
When excited by a short pulse laser,
the emitter can produce a short pulse of an electric field,
which contains a broad frequency component in the THz range.
Although it can produce a high peak amplitude of the electric field,
the average power is limited.
As for a narrow-band source, 
laser-based technologies have been developed.
From high-peak-power optical pulse lasers,
a signal in the THz range can be produced by frequency down-conversion
utilizing a nonlinear crystal.
Although high peak power can be obtained,
these systems are limited to low-duty pulse operation of the pump laser.
Recently, semiconductor laser technology has been under development
in  the THz range. 
A device called the quantum cascaded laser (QCL) can now 
produce milliwatt level of cw power \cite{qcl-kohler, qcl-walther}.
From microwave technology,
a backward wave oscillator (BWO) system
is now commercially available \cite{bwo}.
Basically, it utilizes radiation emitted from the coherent motion of electrons,
and the principle is similar to those developing in accelerator based system.

Accelerator-based sources have been under development.
The conventional ones use incoherent synchrotron radiation in an electron storage ring \cite{nanba}.
When the emission of synchrotron radiation occurs coherently in the structure of a bunch,
very high-power radiation is generated.
A coherent synchrotron radiation (CSR) source can be realized at the storage rings
by using short-bunch beam optics  \cite{wang, circe, inko} or
imprinting electron density modulation by interaction 
with a laser pulse \cite{byrd, uvsorshimada}.
The electron-linac-based system has advantage in producing a short-bunch beam.
When the bunch length is shorter than the radiation wavelength,
the radiation becomes coherent for all the electrons in a bunch.
The radiation intensity is proportional to the square of the bunch charge.
This system can realize a much higher intensity at higher frequencies than ring-based systems,
although the repetition rate is limited by the machine cycle.

Various radiation mechanisms can be considered.
The most conventional one is CSR emitted at a bending magnet \cite{jlabcsr}.
Using an undulator \cite{krafft_source, novo-fel},  
a narrow spectrum of radiation can be produced.
By inserting a target in the path of the beam,
coherent transition radiation (CTR) can be produced at the target.
It is radiation emitted when a relativistic beam passes through an interface 
between materials with different refraction indexes.
The advantages of CTR over CSR are 
that it can be set up in a simple geometry in the straight pass of the beam,
and it has a uniform frequency dependence.
CTR is widely used for electron beam diagnostics of accelerators 
\cite{ctr-murokh, ctr-kung}.
However, it cannot be used as a high-power source
because it destroys the electron beam
and produces background radiation;  also, the target is damaged.
Coherent diffraction radiation (CDR) is a radiation mechanism that is similar to CTR.
Radiation is emitted when an electron beam nondestructively passes
through an aperture or near the edge of a target \cite{dr-potylitsyn}.
It can be used with a high-power beam.
It also has been used for beam diagnostics \cite{odri, karataev}.

Conventional linacs of normal conducting accelerator cavities
only work at low-duty pulsed operation
limited by the rf power loss on the cavity wall.
On the other hand, recently developed superconducting linacs
can be operated in a continuous mode (cw).
The average beam power can be much higher compared with that from normal conducting linacs.
Especially with an energy recovery linac (ERL) scheme \cite{erlreview}
that recycles the beam power,
beam loading effects can be neglected,
and a much higher beam power than that applied by the external rf source
can be operated. 
The features of the high current beam realized by a superconducting linac
fit the nondestructive schemes of THz sources,
such as CSR and CDR,
and it has the potential to be a high-power source \cite{telbe}.

Here, we consider an optical cavity system
that stacks the coherent radiation emitted in the cavity;
in other words,
the beam excites the optical cavity.
When the cavity is excited at resonance,
an electron bunch in a multi-bunch beam coherently emits radiation
in the electromagnetic field that already exists in the cavity.
This results in more radiated power being extracted from the electron bunch
than in a simple setup that is not based on a cavity.
This mechanism is called stimulated radiation.

The principle of stimulated radiation in the THz range
was first tested in a destructive layout by CTR \cite{lihn}.
The electron beam struck one of the mirrors of an optical cavity and emitted CTR.
By measuring the internal cavity power while scanning the cavity length,
a resonance structure was observed.
Tests by CSR were performed 
using a two-mirror cavity with a bending magnet in the center
\cite{shibata_broadband, shibata_ctr, shibata_temporal}.
The test was also performed in a destructive layout,
with the beam striking and being transmitted by one of the mirrors.
Resonance peaks of millimeter waves
were observed with a narrow-band detection setup.
Recently, a test in the millimeter wave range
in a CDR layout that used cavity mirrors with a hole for beam passage 
was performed \cite{aryshev_lucx}.
Although some resonance behavior was observed,
the structure of the resonance peak was not clear.
There might be several reasons for this.
The cavity design was not optimized for broadband excitation.
The number of bunches was limited under the experimental conditions.
There was difficulty in passing the beam without loss.

Broadband stimulated radiation in the THz range in a CDR layout without beam loss
has not been established.
Because it requires a low-emittance short-bunch beam at a high repetition rate,
the experimental setup has become feasible only 
with the development of a recent high-quality linac.
One of the practical problem with CDR layout is
the severe requirement of the passage of a high current beam through a small aperture.
This requires  experimental demonstration \cite{jlab_MW}.
From the standpoint of beam dynamics,
the CDR layout enables us to measure the beam downstream of the cavity.
It provides useful information such as beam deceleration
due to the radiation emission.
Such an energy diagnostic obtained downstream turned out to be quite useful
in free electron laser (FEL) machine studies \cite{lclslh, lclsmbi}.

\begin{figure}[h]
\includegraphics[width=0.9\linewidth]{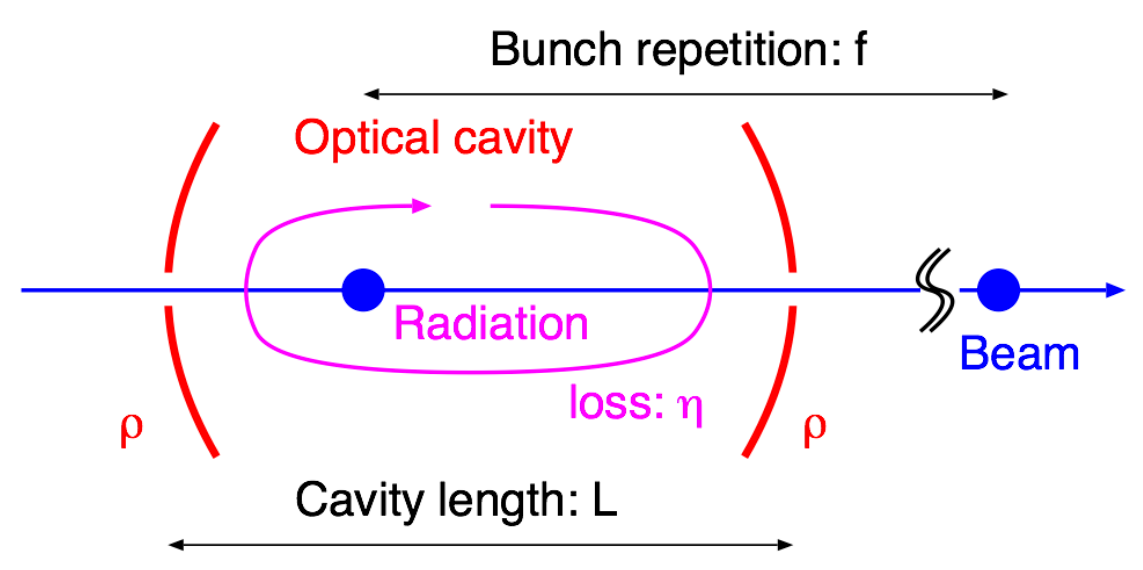}
\caption{\label{fig:scheme} Scheme of the system.
A two-mirror optical cavity is located in a straight pass of the multi-bunch beam.
The beam passes through the center holes of the cavity mirrors.
}
\end{figure}

We present an experiment performed with a modern superconducting linac
constructed as a test facility of an ERL \cite{cerlconstruction}.
The low-emittance beam produced by the photocathode injector 
can realize the CDR layout.
The sub-picosecond short-bunch beam generated 
in the bunch compression mode \cite{cerlbunch}
emits coherent radiation in the THz range.
The schematic of the experimental situation is shown in Figure \ref{fig:scheme}.
An optical cavity formed by two spherical mirrors with a small hole in the center
was installed in a straight pass of an electron beam.
The beam emits radiation in the THz range in the cavity
via the CDR process at the boundary of the mirrors.
When the cavity round-trip time matches the bunch spacing of the multi-bunch beam,
radiation occurs coherently between bunches,
and higher radiation power than just a simple intensity summation of the number of bunches is produced. 
%
%

%
We presented a result showing evidence of stimulated radiation in the THz range
in the CDR layout for the first time 
by measuring the resonance peak while scanning the cavity length \cite{rcdrprl}.
In this paper,
we report detailed studies,
such as
observations of resonance peaks obtained in separated frequencies
and by changing the number of bunches in a beam macro-pulse.
Measurement of the cavity power growth in the time domain was performed
using a fast detector,
and the result was compared with the calculation.
The beam energy was measured downstream of the cavity,
and it showed the effect of beam-to-radiation energy conversion
at the cavity resonance.

\section{Principle}
We describe the principle of excitation of optical cavity modes by beam passage.
Although the mirrors have a hole at the center in the actual setup,
we ignore the effect of the holes in this section.
This situation can be understood as
the case of the CTR cavity, i.e., infinitesimal hole size. 

\subsection{Excitation of an optical cavity mode}
\label{sec:mode}

\begin{figure}[h]
\includegraphics[width=0.8\linewidth]{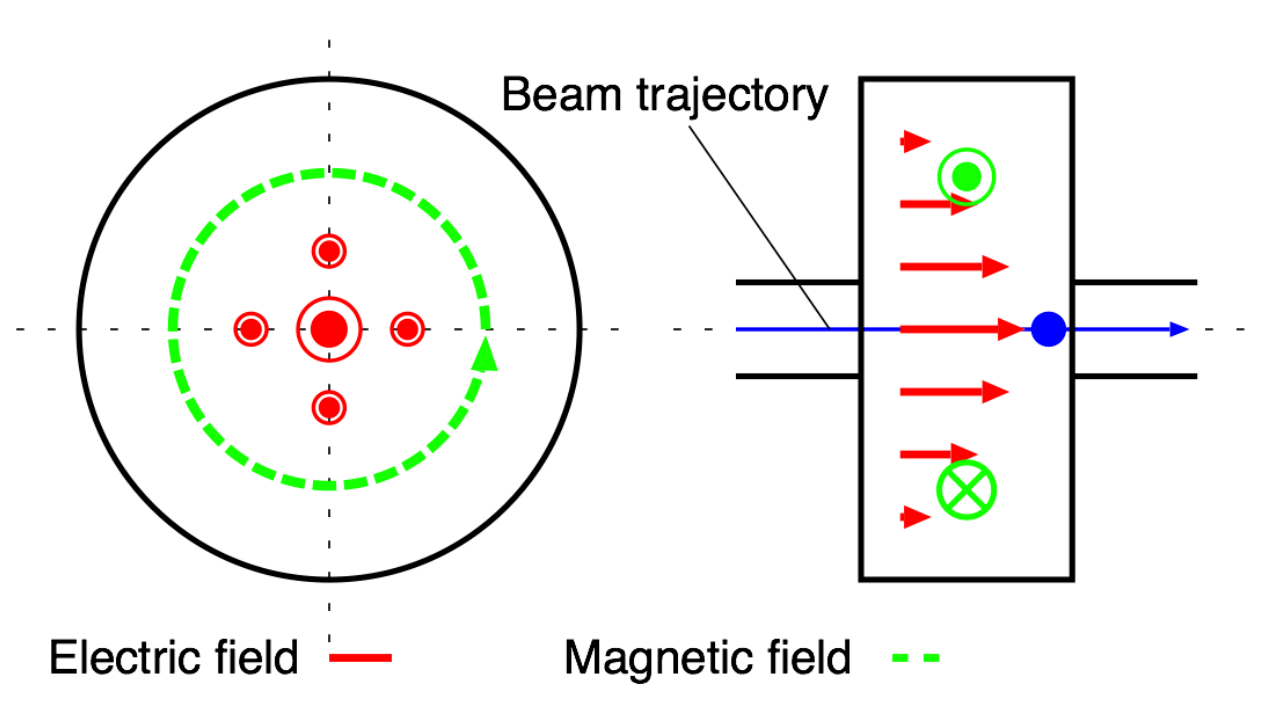}
\caption{\label{fig:cavityinteractionprinciple} Beam and rf cavity interaction.
This shows an example of a cylindrical cavity and its TM010 mode.
The beam works to the mode
via the longitudinal electric field in the beam trajectory. 
}
\end{figure}

First, we recall the conventional procedure for calculating the interaction between
a charged particle and an rf cavity structure in a beam duct,
as illustrated in Figure \ref{fig:cavityinteractionprinciple}.
When a cavity structure is given,
one can calculate its eigen modes. 
The question is
how much electromagnetic power will be excited in a specific eigen mode
by the beam passage \cite{cbpm}.
When an eigen mode of interest is given,
its normalized shunt impedance $(R/Q)$ is defined as
\begin{equation}
R/Q = \frac{|\int \vec{E} d\vec{s}|^2}{\omega U}
\quad ,
\label{eq:roverq}
\end{equation}
where $\omega$ is the angular frequency,
and $\vec{E}$ is the electric field of the mode.
$U$ is the energy stored in the cavity corresponding to the electric field distribution.
The numerator indicates integration of the electric field
along the beam trajectory moving at the speed of the beam.
The excited energy in the cavity can be calculated by
\begin{equation}
U_{exc} = \frac{\omega}{4} (R/Q) q^2
\quad ,
\label{eq:uexc}
\end{equation}
where $q$ is the bunch charge.
Here, we assume that 
the bunch length is negligibly smaller than the wavelength.

%

As can be understood from Eq.~\ref{eq:roverq},
in order to be excited by the beam,
the mode has to have a component of the electric field 
that is longitudinal with respect to the beam trajectory.
In the case of a plane wave that is traveling parallel to the beam, however,
its electric field is perpendicular to the beam trajectory.
Hence, such a mode does not interact with the beam.
In this paper,
we consider an optical cavity that stores optical modes
that travel parallel to the beam.
Because the modes of interest here are not a simple plane wave,
they can be excited by the beam.

By describing the electromagnetic amplitude of the field 
localized in the vicinity of the $z$-axis
and propagating in the $z$ direction
as $E(x,y,z,t) = u \exp(i(\omega t -kz))$,
the time-independent part $u(x,y,z)$
is described by the paraxial equation,
\begin{equation}
i \frac{\partial}{\partial z} u(x,y,z) = 
- \frac{1}{2 k} \left( 
\frac{\partial ^2 }{\partial x^2} + \frac{\partial ^2}{ \partial y^2}
\right) u(x,y,z)
\quad,
\label{eq:paraxial}
\end{equation}
where $x$ and $y$ are the axes transverse to the propagation direction,
and $k$ is the wave number,
which is related to the radiation wavelength $\lambda$
and the speed of light, $c$,
as $k=2 \pi/\lambda = \omega/c$.
The simplest solution of Eq. \ref{eq:paraxial} is
the well-known Gaussian beam
\begin{equation}
u(r,z) = \frac{A w_0}{ w(z)} \exp\left(\frac{-r^2}{w^2(z)} \right)
\exp \left( -i \frac{k r^2}{2 R} + i \phi \right)
\quad ,
\end{equation}
where
$r=\sqrt{x^2 + y^2}$, and
$w_0$ is the beam size at the waist.
$w(z)$ is the beam size at location $z$:
$w(z) = w_0 \sqrt{1 + (z/z_0)^2}$.
$R$ is the curvature radius of the wavefront:
$R(z) = z \left( 1 + (z_0/z)^2 \right)$.
$z_0$ is the Rayleigh length,
which is defined as $z_0 = \pi w_0^2/\lambda$.
$\phi$ is called the Gouy phase:
$\phi(z) = \tan ^{-1} (z/z_0)$.
$A$ is an arbitrary scale factor.

In the case of an optical cavity,
due to the boundary condition of the mirrors,
solutions of discontinuous $\omega$ are accepted as eigen modes
and are called longitudinal modes.
One important difference between an optical cavity
and a typical rf cavity is that
it contains thousands of longitudinal modes.
This results in the picture of mode-lock pulse
that a light pulse propagates back and forth in the cavity.

The fundamental Gaussian mode explained above 
does not have a longitudinal field on the beam axis;
hence, it is not excited by the beam.
Here, we consider higher-order transverse mode
solutions, called the Hermite-Gaussian beam.
The $(l,m)$-th mode can be described as \cite{siegman}
\begin{eqnarray}
u(x,y,z) = \frac{A}{w(z)} H_l (\frac{\sqrt{2}x}{w(z)}) 
 \exp\left(\frac{-x^2}{w^2(z)} \right)  \nonumber \\
 H_m(\frac{\sqrt{2}y}{w(z)})\exp\left(\frac{-y^2}{w^2(z)} \right) \nonumber \\
\exp \left( -i \frac{k r^2}{2 R} + i (l+m+1)\phi \right)
\quad ,
\end{eqnarray}
where $H_n$ is the $n$-th order Hermite function.
For simplicity,
we limit the discussion to a cylindrical symmetric case.
The transverse electric field of the first-order mode is explicitly
written as follows:
\begin{equation}
E^r = \frac{A r}{w^2(z)}
\exp\left(\frac{-r^2}{w^2(z)} \right) 
\exp \left( i(\omega t -kz)  -i \frac{k r^2}{2 R} + i 2 \phi \right)
\quad .
\label{eq:transversefield}
\end{equation}
The round-trip propagating energy in the cavity of length $L$, 
corresponding to $U$ of Eq. \ref{eq:roverq},
can be calculated by integrating the electromagnetic energy as
\begin{eqnarray}
U &=& 2 \times \frac{\epsilon_0}{2} \int |E^r|^2 dV
= 2 \times \frac{\epsilon_0}{2} \int ( |E^x|^2 + |E^y|^2 ) dV
\nonumber \\
&=& \frac{\epsilon_0 \pi A^2}{8} L
\quad ,
\label{eq:energyinteg}
\end{eqnarray}
where $\epsilon_0$ is the vacuum permittivity.
$2 \times$ indicates the contribution of the round-trip.
Here, the contribution of the small longitudinal field component 
which will be discussed in the following, is neglected.

%
\begin{figure}[h]
\includegraphics[width=0.7\linewidth]{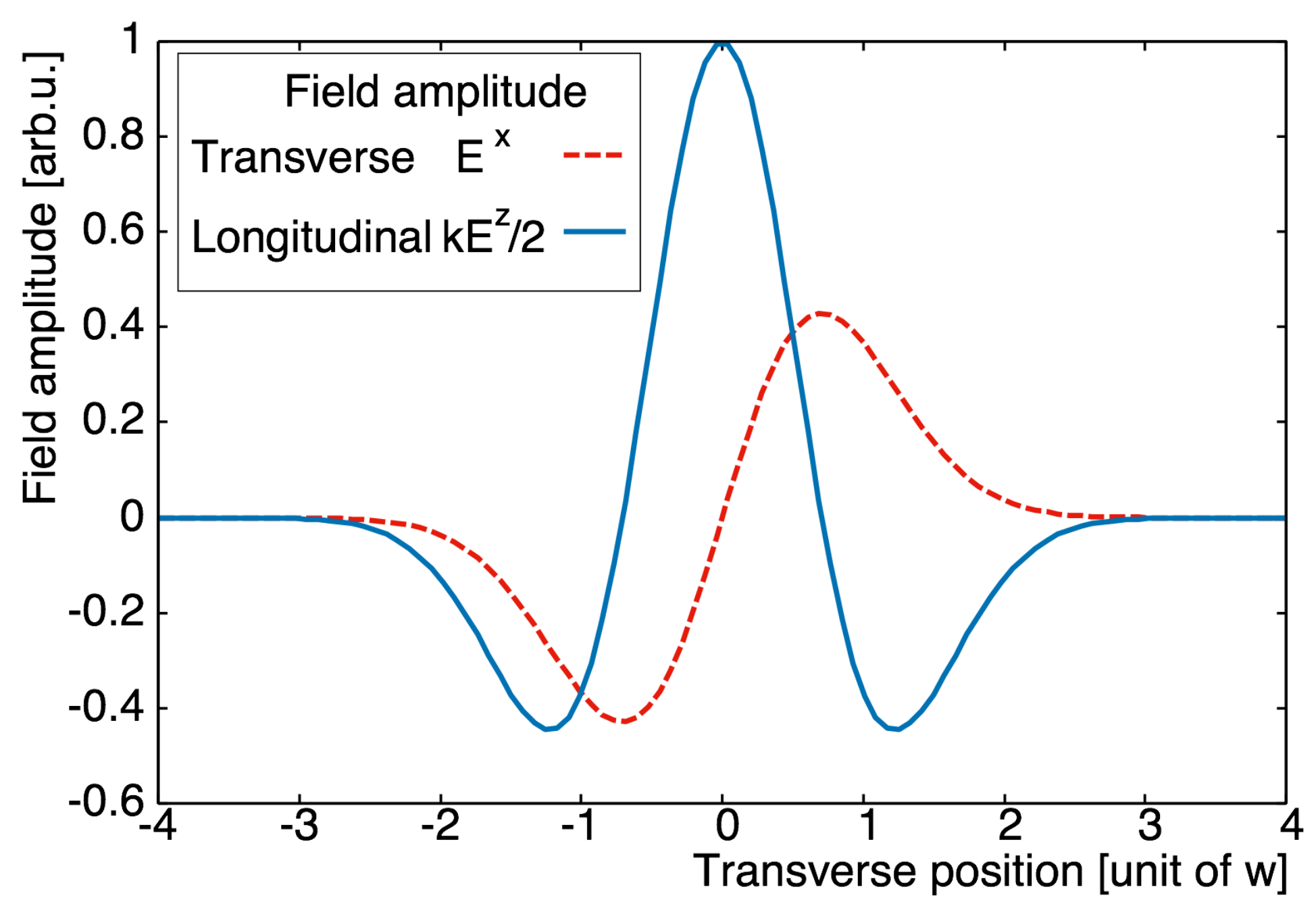}
\caption{\label{fig:lfield} Longitudinal field in the first-order Hermite-Gaussian beam.
At the center of the profile, the longitudinal field component of 
$\sim1/k$ of the transverse peak field is present.
}
\end{figure}

%
Because the three-dimensional field components are related to each other,
the longitudinal component of the electric field can be determined by
the transverse field distribution \cite{laseracc}.
The general relation that connects the longitudinal and transverse fields is given as
\begin{equation}
i k E^z = \frac{\partial E^x}{\partial x} +  \frac{\partial E^y}{\partial y}
 = 2 \frac{\partial E^r}{\partial r}
\label{eq:ez}
\quad .
\end{equation}
Applying Eq. \ref{eq:ez} to
Eq. \ref{eq:transversefield},
the longitudinal field distribution can be calculated
as shown in Figure \ref{fig:lfield}.
At the center of the beam,
the longitudinal field is maximum,
while the transverse field is zero.
The longitudinal field on the axis can be explicitly written as
\begin{equation}
E^z = -\frac{A}{k w^2(z)} \exp( i(\omega t -kz) + i 2 \phi(z))
\quad .
\end{equation}
Approximating the velocity of the beam as the speed of light,
$\omega t - kz =0$,
the excited energy becomes 
\begin{eqnarray}
U_{exc} &=& \frac{q^2}{4} \frac{|E^z ds|^2}{U} \\
&=& \frac{q^2}{ \epsilon_0 \pi L}
\left|
\int \frac{\exp\left(  i 2 \tan^{-1}(z/z_0) \right)}{z_0 (1+(z/z_0)^2)}dz
\right|^2
\quad .
\end{eqnarray}
The integration along the $z$-axis should be performed between the cavity mirrors.
As can be seen,
the result basically depends only on
$z_0$ and the cavity boundaries.

\begin{figure}[h]
\includegraphics[width=0.8\linewidth]{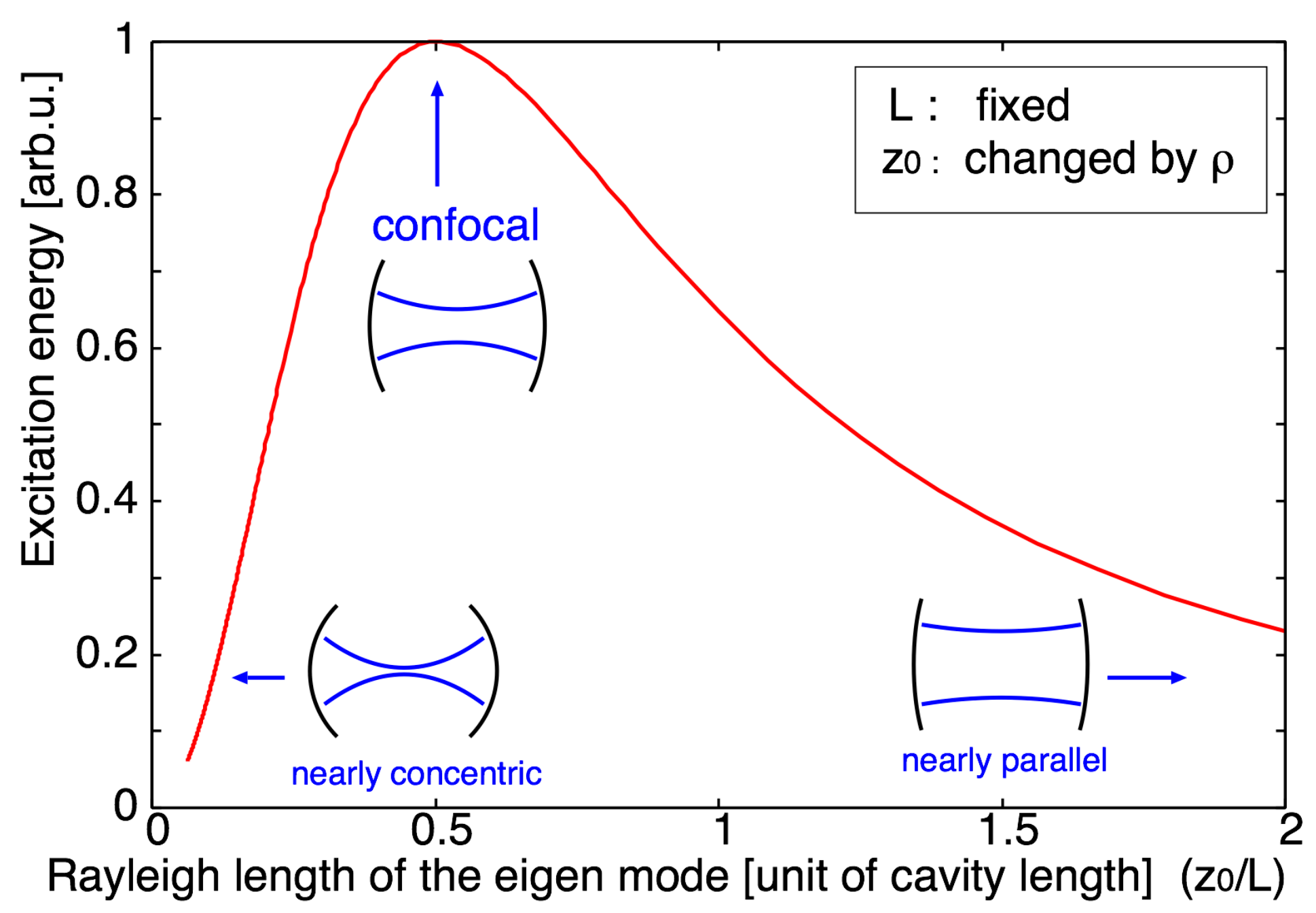}
\caption{\label{fig:cavparameter} Dependence of beam to cavity coupling on the cavity parameter.
The Rayleigh length of the mode $z_0$ can be controlled by the mirror curvature $\rho$.
The types of cavity change from parallel, confocal,  to concentric.
The coupling is maximum at the confocal cavity in the case of the symmetric cavity.
}
\end{figure}

$z_0$ depends on the design of the cavity,
which is determined by the cavity length $L$ and the curvature radius of the mirrors $\rho$.
Figure \ref{fig:cavparameter} shows
that the relative beam to cavity coupling strength,
in other words, 
the efficiency of cavity mode excitation, depends on the cavity parameter
under a fixed cavity length
for the symmetric cavity case.
The highest coupling is realized at $2 z_0= L$.
It corresponds to $L=\rho$,
and this type of cavity is called confocal.

%
%

\subsection{Excitation by a multi-bunch beam}
\label{sec:mbexcitation}

\begin{figure}[h]
\includegraphics[width=0.8\linewidth]{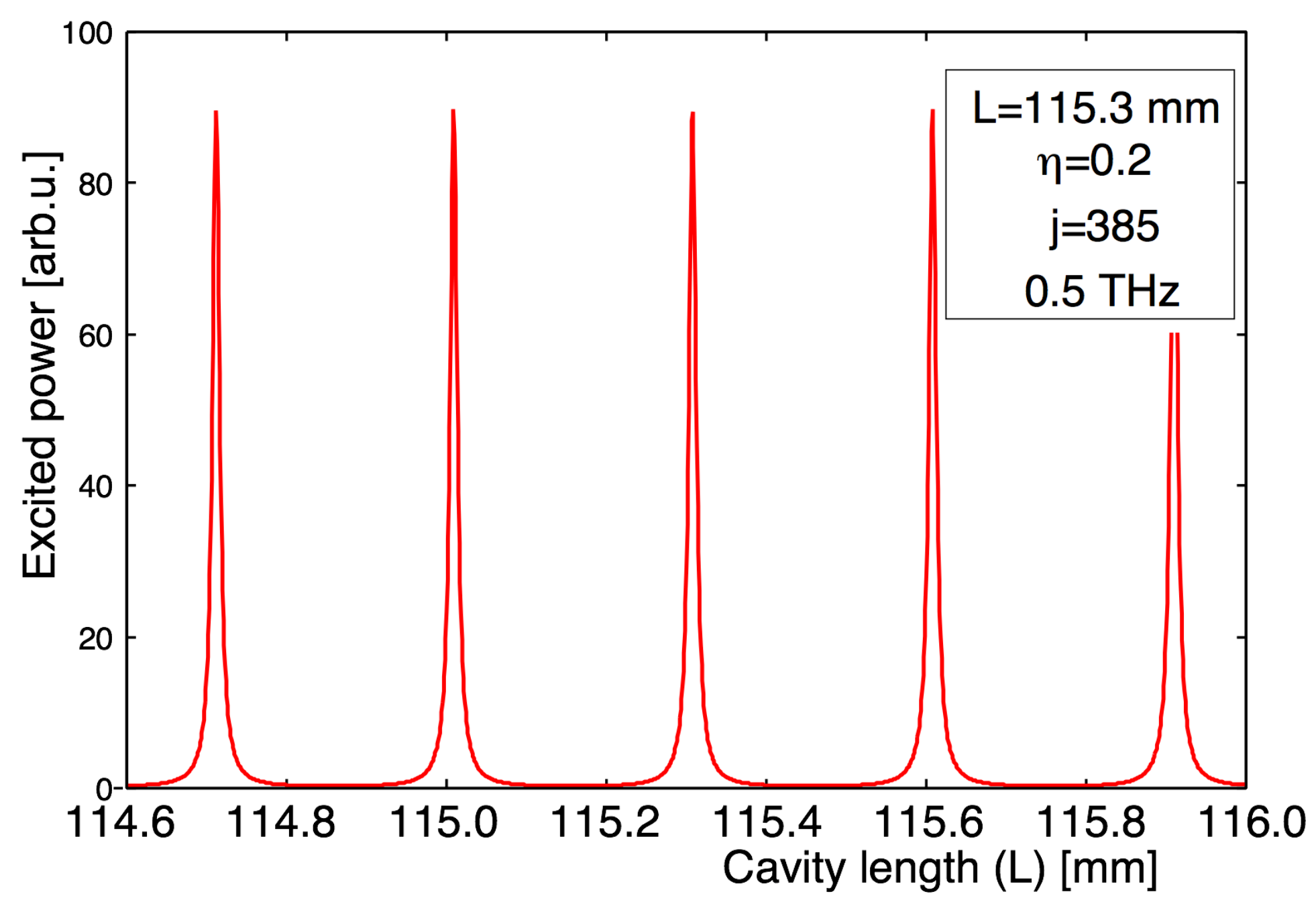}
\caption{\label{fig:cavityresonance} Excitation of a mode by a multi-bunch beam.
For a mode of wavelength $\lambda$,
sharp resonance peaks are realized every $\lambda/2$ of the cavity length.
The width of the resonance is determined by the cavity loss.
}
\end{figure}

When the cavity is excited by a multi-bunch beam 
whose bunch spacing matches the round-trip time of the cavity,
the radiation signal of each bunch is stacked as a coherent amplitude addition.
The amplitude of a mode in the cavity after the $n$-th bunch becomes
\begin{equation}
v_n = v_1 \sum_{m=1}^n (\sqrt{1- \eta} e^{i \theta})^m
\quad ,
\end{equation}
where $\eta$ is the relative power loss in one round trip;
$\eta=1$ indicates the total loss, and $0$ the zero loss.
$\theta$ is the phase shift over a single round trip.
$\theta$ can be changed by finely changing $L$
in a scale within the wavelength.
The power enhancement gain after an infinite number of bunches
can be obtained as
\begin{equation}
G = \frac{|v_{\infty}|^2}{|v_1|^2}
= \frac{1}{2 - \eta -2\sqrt{1-\eta}\cos \theta}
\quad ,
\end{equation}
and is plotted in Figure \ref{fig:cavityresonance}
as a function of $L$.
We note that this resonance function is the same as the one that appears 
when an external optical cavity is excited by a laser \cite{siegman}.
We define the finesse of the cavity by $F=2 \pi / \eta$.
The gain at resonance is $G = 4/\eta^2$.
The cavity mode is excited more strongly at a lower cavity loss.
When we are interested in
the extracted radiation power
available outside the cavity, instead of the cavity internal power,
the effective power gain becomes
$ G_{eff} \sim \eta G $.
Here, we assume that the cavity loss mostly corresponds to the out-coupling.

%
The full-width-half-maximum resonance width is determined by the cavity loss.
It can be written using the cavity length variation as
\begin{equation}
\Delta L \sim \frac{\lambda}{2F} = \frac{\lambda \eta}{4 \pi}
\quad .
\end{equation}
Further, it can be written using the bunch timing variation as
\begin{equation}
\Delta T \sim \frac{\lambda}{2 c F}
\quad .
\end{equation}
In order to resonantly excite a cavity of small loss,
mechanical stability of the cavity length and
stability of the bunch arrival time are required.
For example, 
in the case of $\lambda=300$ $\mu$m (1 THz) and $F=1000$ ($\eta \sim 0.006$),
$\Delta L = 150$ nm and $\Delta T = 0.5$ fs.

\subsection{Broad-band excitation}
\label{sec:broadband}
The discussion in Sec.\ref{sec:mbexcitation} focused on one of the cavity modes.
As explained in Sec.\ref{sec:mode},
the optical cavity has many longitudinal modes
(and sets of transverse modes belonging to each longitudinal mode)
that correspond to frequencies that are integral multiples of the round-trip frequency.
The frequency of the $j$-th longitudinal mode
is $f j$, where $f$ is the fundamental frequency and matches the bunch repetition.
Here, we consider the shift of resonance conditions in the longitudinal modes.
For simplicity, we limit the discussion to the symmetric cavity and the first-order transverse mode.
The additional phase effect due to the Gouy phase 
affects the resonance condition of the cavity by 
$
2(2 \phi(L/2) - 2 \phi(-L/2))
$
as the round-trip phase.
This shifts the resonance condition
for the $j$-th longitudinal mode as
\begin{equation}
\Delta \theta ^{(j)}
= 2 \pi \left(
j - \frac{4}{\pi} \tan^{-1} \sqrt{\frac{L/\rho}{2-L/\rho}}
\right)
\quad ,
\end{equation}
where $\rho$ is the curvature radius of the mirrors.

\begin{figure}[h]
\includegraphics[width=0.95\linewidth]{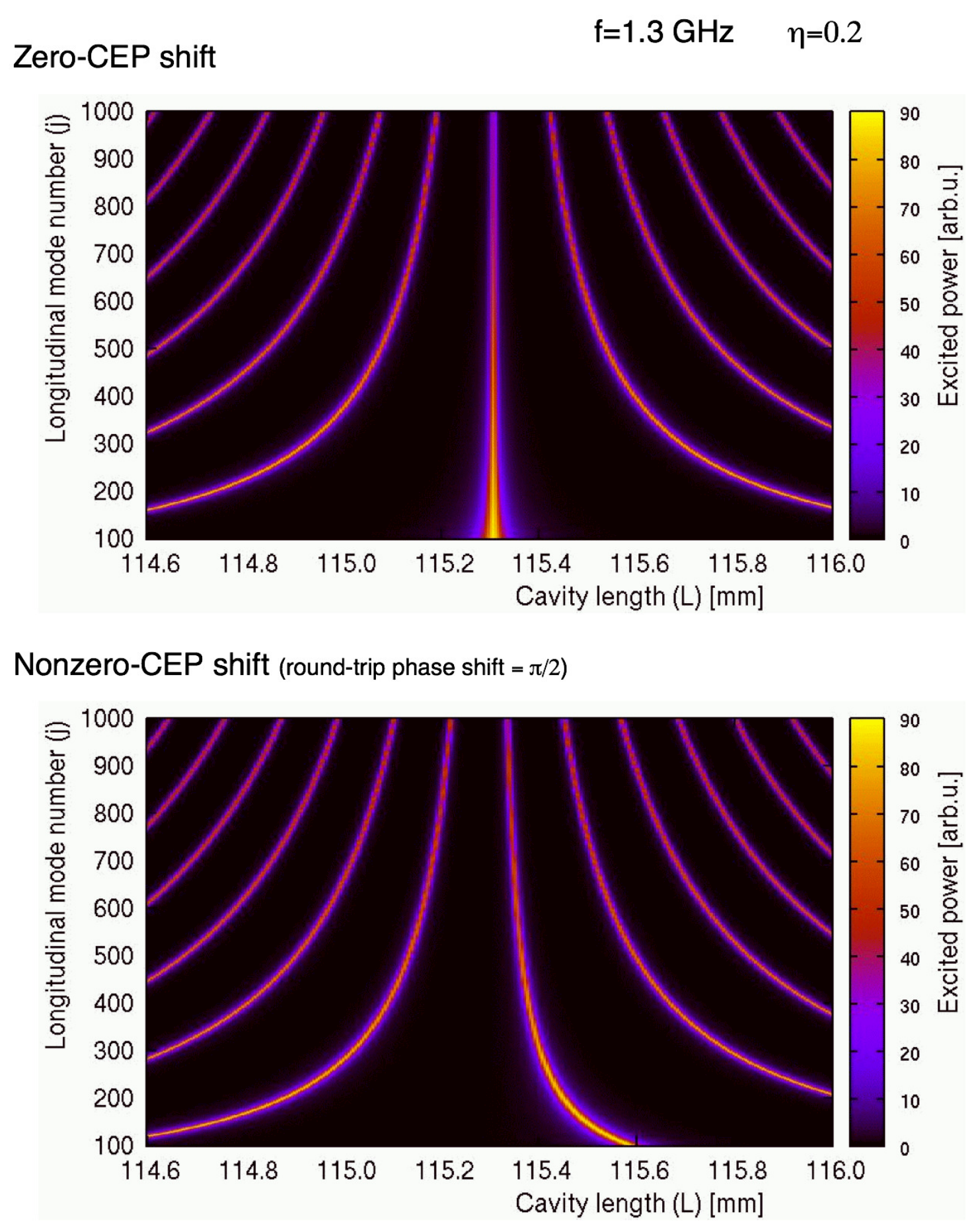}
\caption{\label{fig:mapping} Example calculation of cavity resonance
on a map of cavity length and the longitudinal mode number.
(Top) Cavity designed to be zero-CEP shift.
The resonance conditions of all the modes
coincide under the perfect synchronization condition,
$L=115.3$ mm in this case.
(Bottom) Cavity designed to be nonzero-CEP shift.
There is no special condition to excite broadband of the modes at the same time.
}
\end{figure}

\begin{figure}[h]
\includegraphics[width=0.9\linewidth]{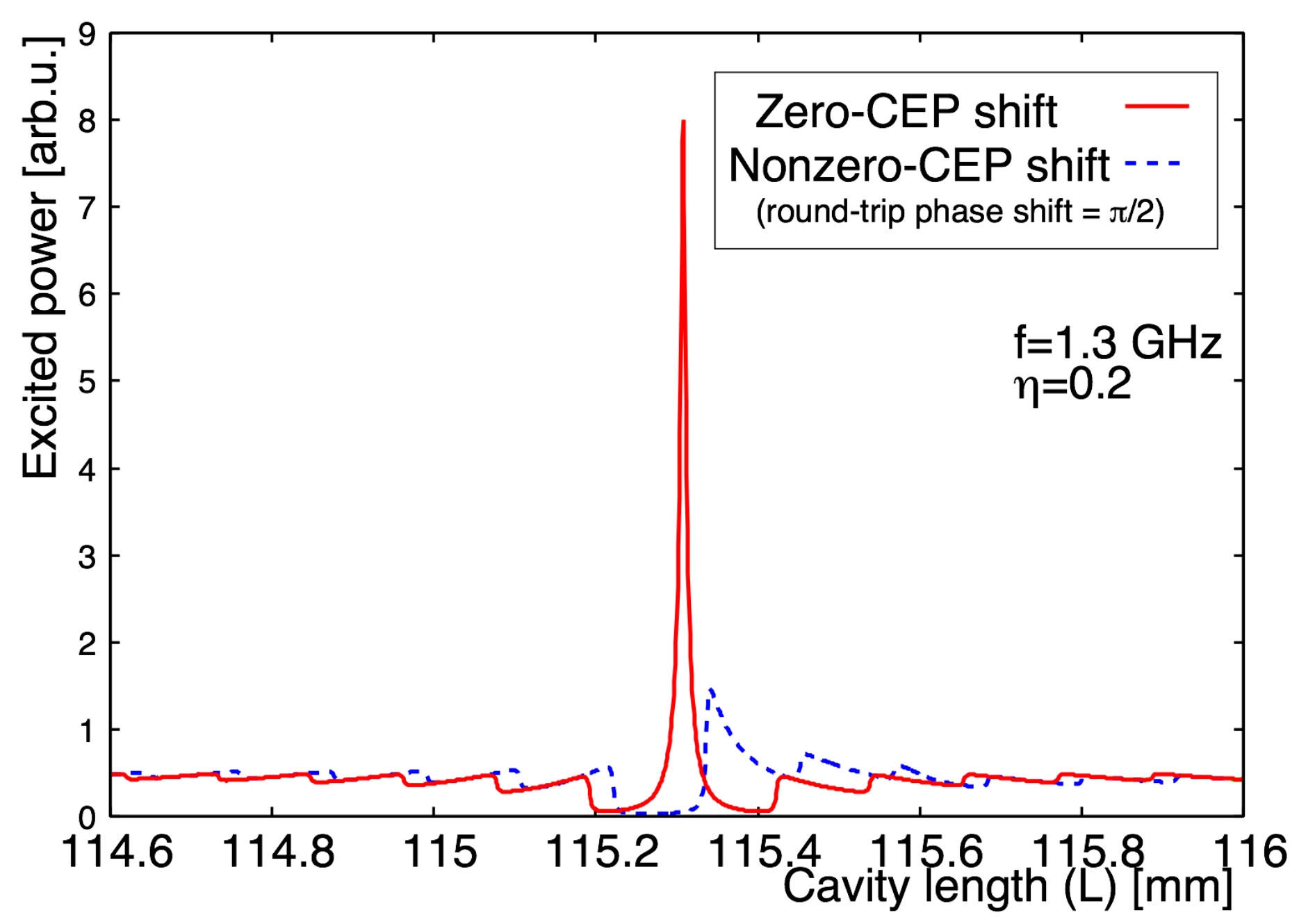}
\caption{\label{fig:scansim} Simulation of the excited radiation power in cavity length scan.
This is given as the projection of the contribution of all the modes in Figure \ref{fig:mapping}.
If the cavity is designed to be zero-CEP shift,
a strong peak appears at the cavity length of perfect synchronization.
}
\end{figure}

In general,
the resonance condition in the cavity length is different for each mode.
Hence, modes of different frequency
cannot be excited by a multi-bunch beam at the same time.
However, in the special case of confocal cavity design, $L/\rho =1$,
$\Delta \theta^{(j)}$ becomes an integral multiple of $2\pi$ for all $j$'s,
resulting in all longitudinal modes being excited simultaneously.
This mechanism can be described
as a picture of the carrier-envelope phase (CEP)
of the pulse traveling back and forth in the optical cavity.
In the general case, the CEP shifts in every round trip of the pulse.
Signals of the multi-bunch beam 
cannot be stacked coherently in such a case. 
The confocal cavity is a special case with a zero-CEP shift,
and hence it can coherently add broad spectral signal in a multi-bunch.
Figure \ref{fig:mapping} explains the cavity resonance 
on a map of the cavity length and longitudinal mode number.
If one only focuses on the longitudinal modes in a narrow bandwidth,
the resonance structure repeats at every half wavelength 
in the cavity length, as already seen in Figure \ref{fig:cavityresonance}.
But in a broad bandwidth,
the resonance condition spreads
due to the wavelength difference.
Only under the perfect synchronized condition,
in which means that the cavity length exactly matches the bunch repetition,
do all the longitudinal modes resonate at the same time.
Note that the broadband excitation is possible
only with the zero-CEP shift design.
The bottom figure of Figure \ref{fig:mapping}
shows an example of a nonzero-CEP shift case,
i.e., with $\rho$ chosen to be non-confocal,
where the resonance line curved as the mode number;
broadband excitation cannot be realized.
When one observes the radiation with a broadband detector 
while scanning the cavity length,
the projected resonance curve
as shown in Figure \ref{fig:scansim} will be seen.
If the cavity is designed to be zero-CEP shift,
a strong peak 
that corresponds to the excitation of the broad spectrum
is seen under the perfect synchronized condition.
The other peaks are smeared out and appear as a plateau.

%
When broadband excitation is realized,
the total radiation power can be estimated 
by considering the following factors:
excitation of a mode by a single bunch,
the effective power gain of stimulated radiation by a multi-bunch,
and the number of longitudinal modes excited at the same time.
The upper limit of the radiation frequency, $\xi _c$,
is given by the bunch length of the electron beam.
The total radiation power becomes,
\begin{equation}
P_{total} =
\frac{U_{exc}}{2L/c} \times G_{eff} \times \frac{\xi_c}{c/2L}
= \frac{4 \xi_c q^2}{\eta \pi \epsilon_0 L}
\quad .
\end{equation}
Because of the energy conservation,
the electron beam loses its kinetic energy
corresponding to the energy converted to the radiation.

\section{Experimental Setup}
\subsection{Accelerator}

\begin{figure}[h]
\includegraphics[width=1.0\linewidth]{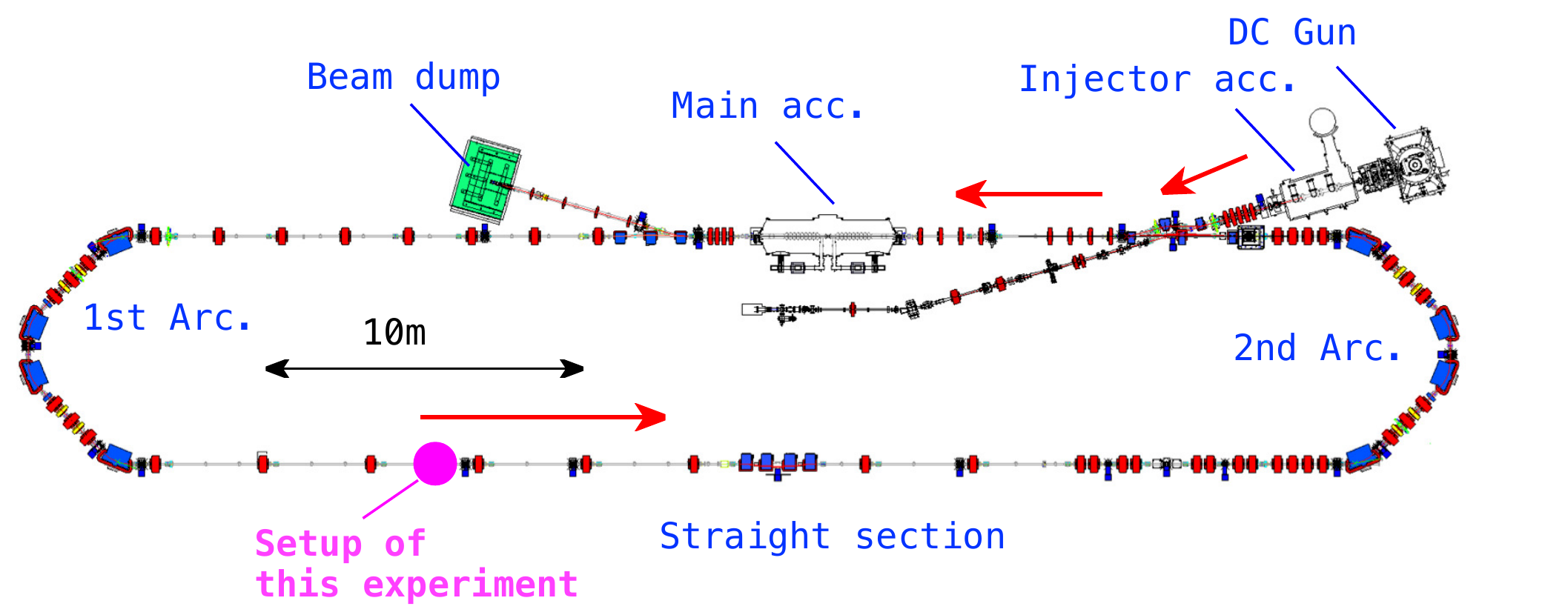}
\caption{\label{fig:cerlfig} Layout of cERL.
The experimental setup is installed 
at the upstream area of the straight section in the return loop.
}
\end{figure}

We performed the experiment with a compact ERL (cERL) in KEK.
cERL is a test facility for the development of 
accelerator technologies for an ERL.
A detailed description of cERL is given elsewhere \cite{cerlconstruction}.
Figure \ref{fig:cerlfig} shows the layout of the cERL.
The beam is produced at the photocathode DC gun.
After it is accelerated by the injector superconducting accelerator up to 
a total energy of 5.3 MeV,
it is merged with the circulating path.
The beam is then accelerated by the main superconducting accelerator
up to a total energy of 17.9 MeV.
After passing through the return loop,
which consists of two arc sections and a straight section,
the beam re-enters the main accelerator in the deceleration phase
and it loses energy down to the injector energy.
It is then sent to the beam dump.
Because the main accelerator recycles the energy by decelerating the beam,
it does not have a net beam loading and can operate with high beam power.　

%
Although the nominal operation mode of cERL
is the energy recovely mode,
this experiment was performed in the burst operation of the energy nonrecovery mode.
The beam was stopped at the end of the return loop.
In the burst operation,
the electron beam emission at the gun
was time-gated by the photocathode laser system
to form a macro-pulse of a multi-bunch beam,
while all the rf systems in the accelerator cavities operated in the cw mode.
The bunch repetition rate was 1.3 GHz,
which is determined by the rf frequency of the accelerator cavities.
The beam conditions in this experiment are summarized in Table \ref{tab:beamparam}.

\begin{table}[hb]
\caption{\label{tab:beamparam}
Beam parameters in this experiment.}
\begin{ruledtabular}
\begin{tabular}{ll}
Beam energy at return loop & 17.9 MeV  \\
Bunch repetition & 1.3 GHz \\
Bunch charge & 1.2 pC \\
Macro-pulse length & 1 $\mu$s (can be changed)\\
Repetition of macro-pulse & 5 Hz\\
Normalized emittance & 2.0 (X), 0.7 (Y) $\mu$m\\
RMS bunch length & 120 fs\\
\end{tabular}
\end{ruledtabular}
\end{table}

\begin{figure}[h]
\includegraphics[width=1.0\linewidth]{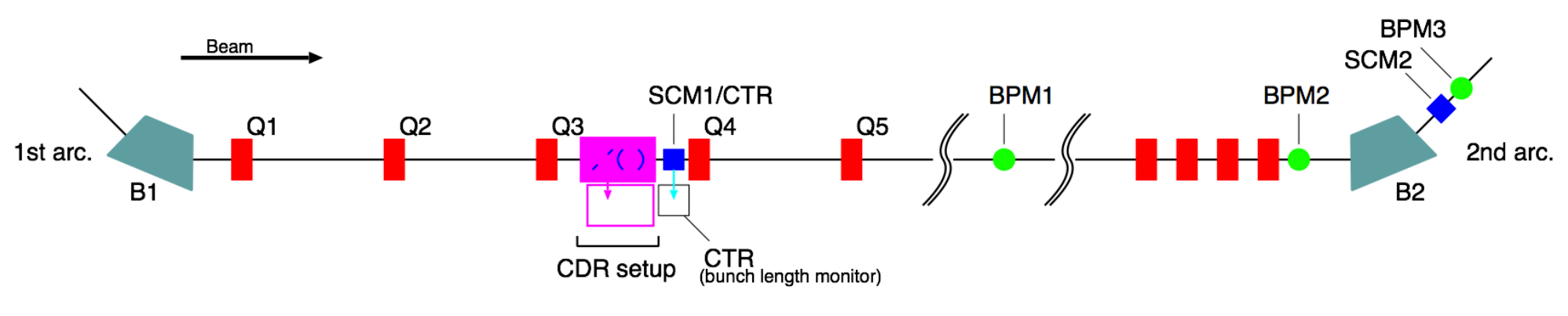}
\caption{\label{fig:straightsectionlayout} Layout of the straight section.
B$i$ are bending magnets, Q$i$ are quadrupole magnets,
BPM$i$ are beam position monitors,
and SCM$i$ are screen monitors. Here $i$ is an integer.
}
\end{figure}

The experiment is set up at the straight section,
where a short-bunch beam is available.
The detailed layout of this section is shown in Figure \ref{fig:straightsectionlayout}.
Because of the achromatic optics of the arc section,
transverse dispersion is designed to be zero in the straight section.
The first three quadrupole magnets in the straight section can be used 
to adjust the beta function of the beam for this experiment.
The quadrupole magnets have correction coils to produce a dipole field 
for finely steering the beam orbit.
The screen monitor system just behind the setup of this experiment
can be switched to a metal target for producing CTR.
It was used for bunch length diagnostics.
The first screen monitor in the second arc section
was used to measure the beam energy variation.

%

\subsection{Optical cavity}

\begin{figure}[h]
\includegraphics[width=0.9\linewidth]{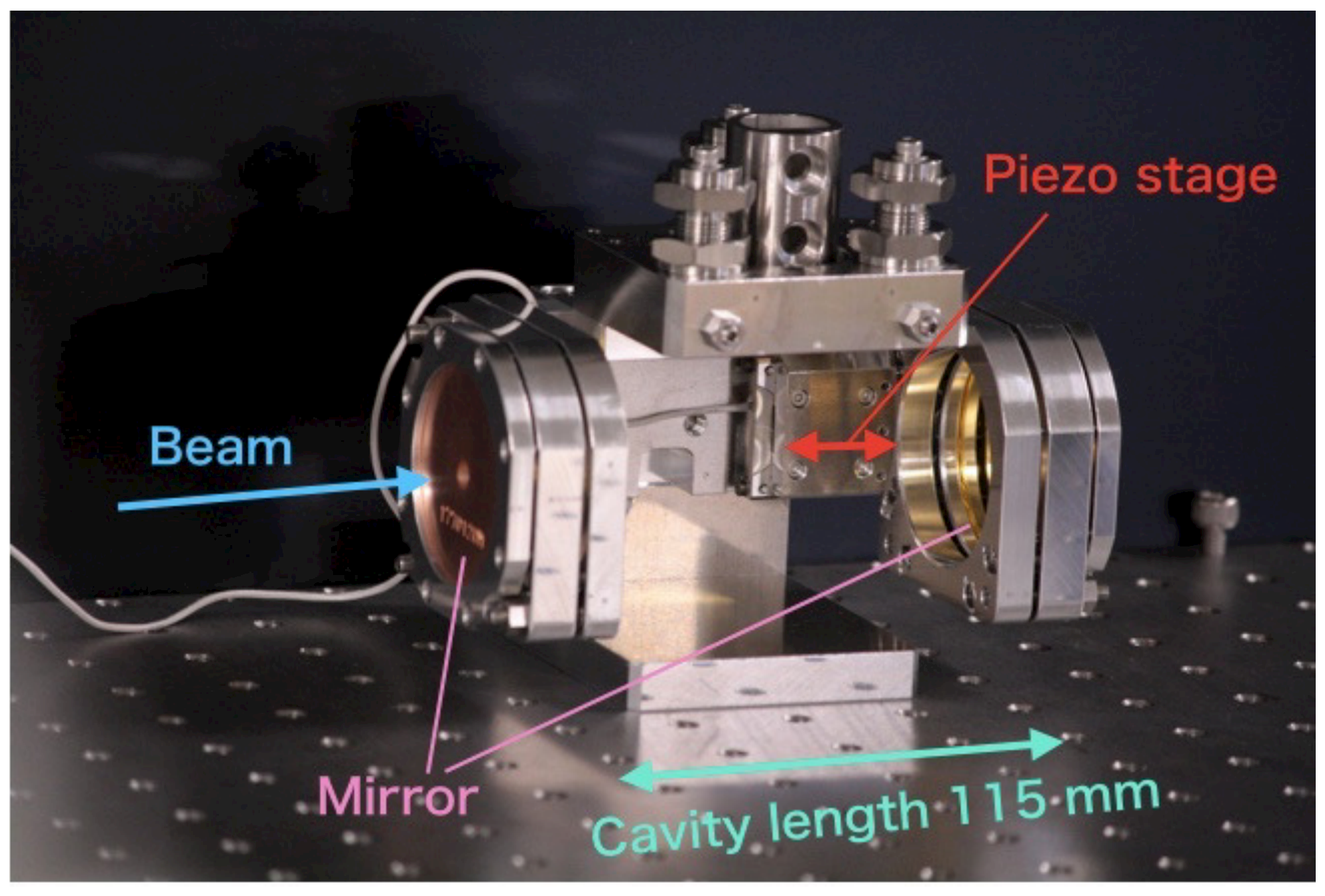}
\caption{\label{fig:cavitypic} Picture of the optical cavity.
}
\end{figure}

Figure \ref{fig:cavitypic} shows the structure of the optical cavity,
which consists of two spherical mirrors facing each other.
The distance between the two mirrors, the cavity length, 
is designed to be 115 mm.
The round-trip time of the cavity matches to the bunch repetition rate of 1.3 GHz.
The two cavity mirrors are identical.
The mirror substrate is made of copper.
Its spherical surface is gold coated to obtain a high reflectance in the THz range.
The thickness of the mirrors is 10 mm,
and their diameter is 50 mm.
The radius of curvature of the mirror is specified to be 115 $\pm$ 3 mm,
which is equal to the cavity length for confocal cavity design.
The opposite surface of the mirror is flat.
The mirror has a hole at the center.
The diameter of the hole is 3 mm at the spherical surface,
and it is enlarged to 6 mm at the opposite surface
having a tapered shape.
Because the mirror substrate was fabricated by rotation machining
of better than 20 $\mu$m precision,
the outer shape, i.e., the flat surface and the cylindrical surface,
can be used as a good mechanical reference
in the initial alignment.
The mirror holders have a kinematic adjustment system
to finely align the angle of the mirror.
In order to scan the cavity length,
the downstream mirror holder is mounted on a piezoelectric stage 
(Model 3030 of attocube Inc.).
It can be controlled up to 1 nm precision in a closed loop system
within a $\pm$10 mm range.

\begin{table}[hb]
\caption{\label{tab:cavityparam}
Parameters of the optical cavity.}
\begin{ruledtabular}
\begin{tabular}{ll}
Cavity length ($L$) & 115 mm  \\
Range of $L$ adjustment & $\pm$10 mm  \\
Diameter of mirror & 50 mm \\
Thickness of mirror & 10 mm \\
Radius of curvature of mirror ($\rho$) & 115 $\pm$ 3 mm\\
Diameter of mirror hole & 3 mm (tapered to 6 mm)\\
Material of mirror & Au-coated Cu \\
\end{tabular}
\end{ruledtabular}
\end{table}

\begin{figure}[h]
\includegraphics[width=0.9\linewidth]{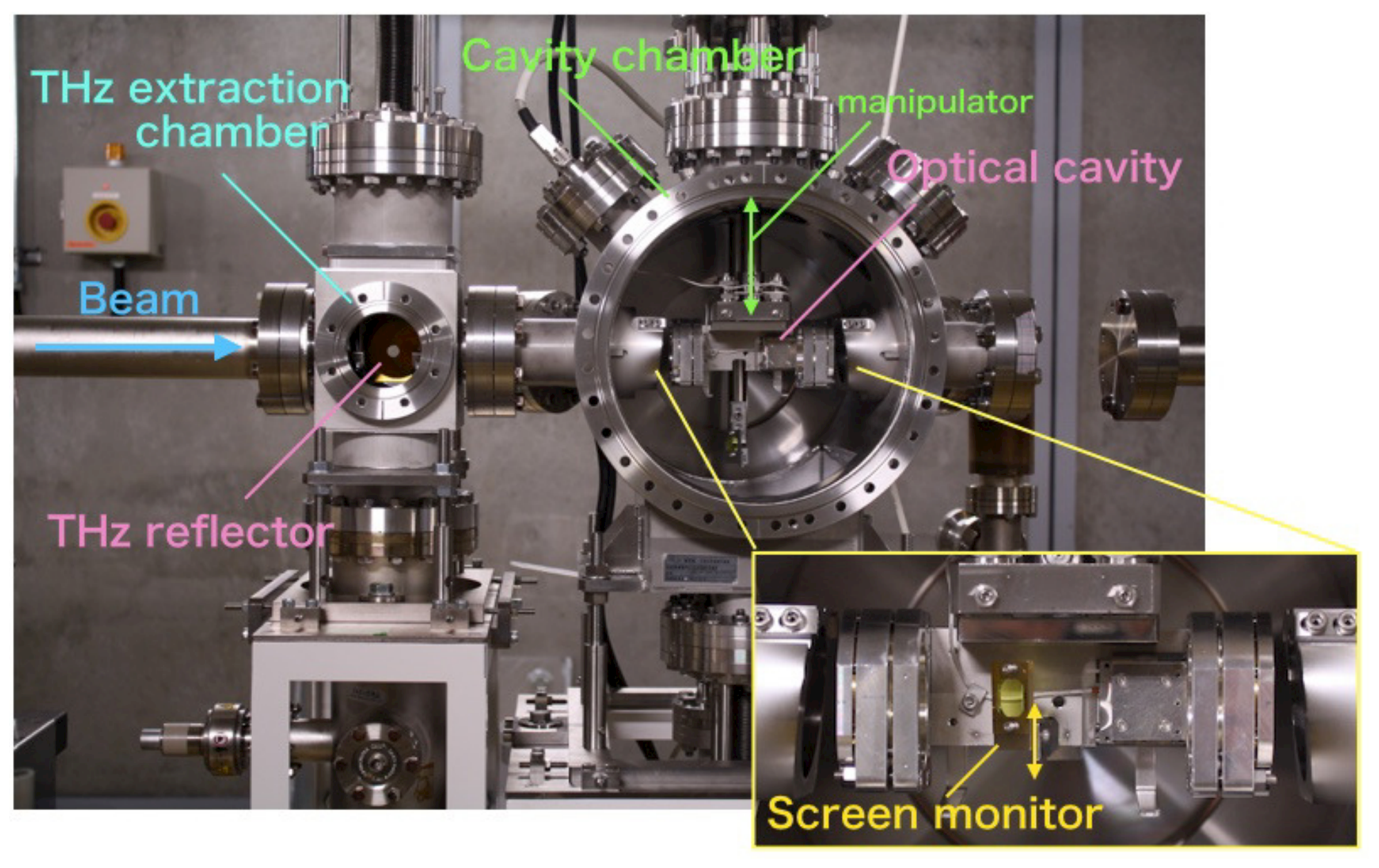}
\caption{\label{fig:chambersetup} Setup of the vacuum chamber.
The cavity is suspended from the top manipulator.
A separated chamber is connected 
next to the cavity chamber to extract the THz radiation.
}
\end{figure}

As shown in Figure \ref{fig:chambersetup},
the cavity is installed in a big cylindrical vacuum chamber
of 244 mm diameter.
The cavity structure is suspended by a manipulator
so that the structure can be repositioned upward from the beam axis in the chamber.
A shielding duct to facilitate smooth connection of
the beam duct can be inserted in the space.
This is to protect against interference with the accelerator operation from other experiments.
A screen monitor consisting of a Ce-doped YAG scintillator of 12$\times$12 mm$^2$
and 0.1 mm thickness
can be inserted at the center of the cavity
at a 45$^{\circ}$ angle with respect to the beam axis.
This screen monitor was used for measuring the beam size
and position at the cavity.
It could also be used for 
blocking the cavity resonance to confirm the resonance signal.

%

%

The optical cavity was designed to be symmetric and confocal.
The eigen mode beam size at the center $w_0$, 
and on the mirrors $w_1$
are given as
\begin{equation}
w_1 = \sqrt{2} w_0 = \sqrt{\frac{L \lambda}{\pi}} \quad .
\end{equation} 
In our setup, $w_1 = 4.7$ mm for 0.5 THz radiation.

Because the first-order transverse mode has a node at the center,
the power loss at the hole can be small enough 
even with a hole of a few millimeters size.
The contribution of the power
in the central 3-mm-diameter area
is estimated to be 1\% of the total power of the mode for one mirror at 0.5 THz.

A fraction of the radiation power inside the cavity 
is emitted through the mirror holes in both directions.
The tapered structure of the mirror hole
was optimized to function as a horn antenna
to improve the directivity of the radiation emission \cite{hornantenna}.
Only emission in the upstream direction is measured in this experimental setup.
As seen in Figure \ref{fig:chambersetup},
a separated vacuum chamber is connected 
309 mm upstream of the cavity chamber.
It works as a radiation extraction chamber
that reflects the radiation from downstream
to the transverse direction by a gold-coated flat stainless steel
plate mirror angled 45$^\circ$ with respect to the beam line.
It has an elliptic hole, which appears as a 10-mm-diameter circular aperture
from the perspective of the electron beam, on the beam axis.
The radiation is transmitted to air 
through a sapphire window of 1.5 mm thickness.

\begin{figure}[h]
\includegraphics[width=0.95\linewidth]{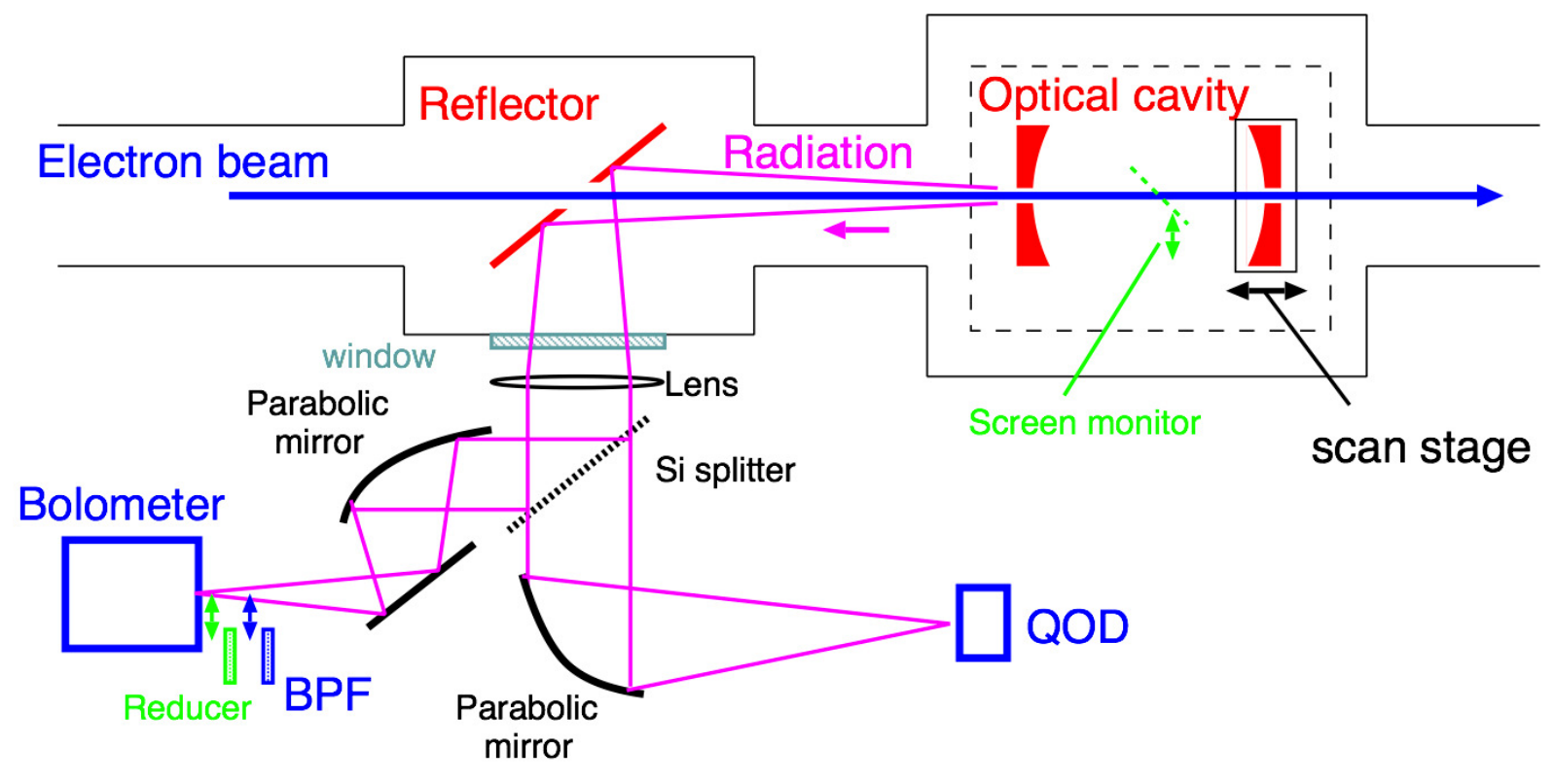}
\caption{\label{fig:detectorlayout} Setup of the THz detectors.
the THz radiation extracted from the chamber
is focused by parabolic mirrors 
and is measured by two types of detectors,
a bolometer and a QOD.
}
\end{figure}

The layout of the THz detection system is shown in Figure \ref{fig:detectorlayout}.
The radiation extracted from the vacuum chamber is 
first collimated by a plastic (PFTE) lens of 500 mm focal length.
It is then split into two paths by a Si plate beam splitter.
One of the paths is for detection with 
a liquid-helium cooled Si bolometer (product of Infrared Laboratories Inc.).
The bolometer has a sensitivity up to 20 THz.
Its time resolution is $\sim$100 $\mu$s,
which is not enough to observe the time structure of power growth.
Reflecting with an off-axis parabolic mirror of focusing length 230 mm,
the radiation is focused on the detector aperture.
We prepared a remote-controlled filter mount in front of the detector.
Four types of bandpass filters (BPFs),
with center frequencies of 0.3, 0.5, 1.0, and 1.5 THz
of 10\% bandwidth (full width at half maximum),
can be inserted.
If the detector is saturated with too large signal,
three types of doped-Si power reducers, with transmittances of 10, 3, and 1\%,
can be inserted in front of the detector.
The second path is for detection with a diode-type detector
that has a time resolution of a few nanoseconds,
which is sufficient to resolve the time structure in a beam macro-pulse. 
At the focal point of an  off-axis parabolic mirror of focusing length 230 mm,
a quasi-optical detector (QOD) (product of Virginia Diodes Inc.)
is mounted on a two-dimensional stage.
The sensitive frequency range of the QOD is lower than that of the bolometer and is not uniform.
It mainly measures a lower frequency range than 0.3 THz.

%

\subsection{Beam tuning}

For realizing this experimental layout of a small beam aperture without serious beam loss,
low beam emittance is the key.
We measured the beam emittance in the straight section
under the same beam conditions
as in this experiment,
i.e.,
a bunch charge of 1.2 pC and
in bunch compression mode.
The measurement was performed with a waist-scan method.
A quadrupole magnet and a screen monitor, 
namely, Q3 and SCM1 in Figure \ref{fig:straightsectionlayout}, were used.
The normalized emittance was estimated to be 2.0 $\mu$m (horizontal)
and 0.7 $\mu$m (vertical), respectively.

\begin{figure}[h]
\includegraphics[width=0.8\linewidth]{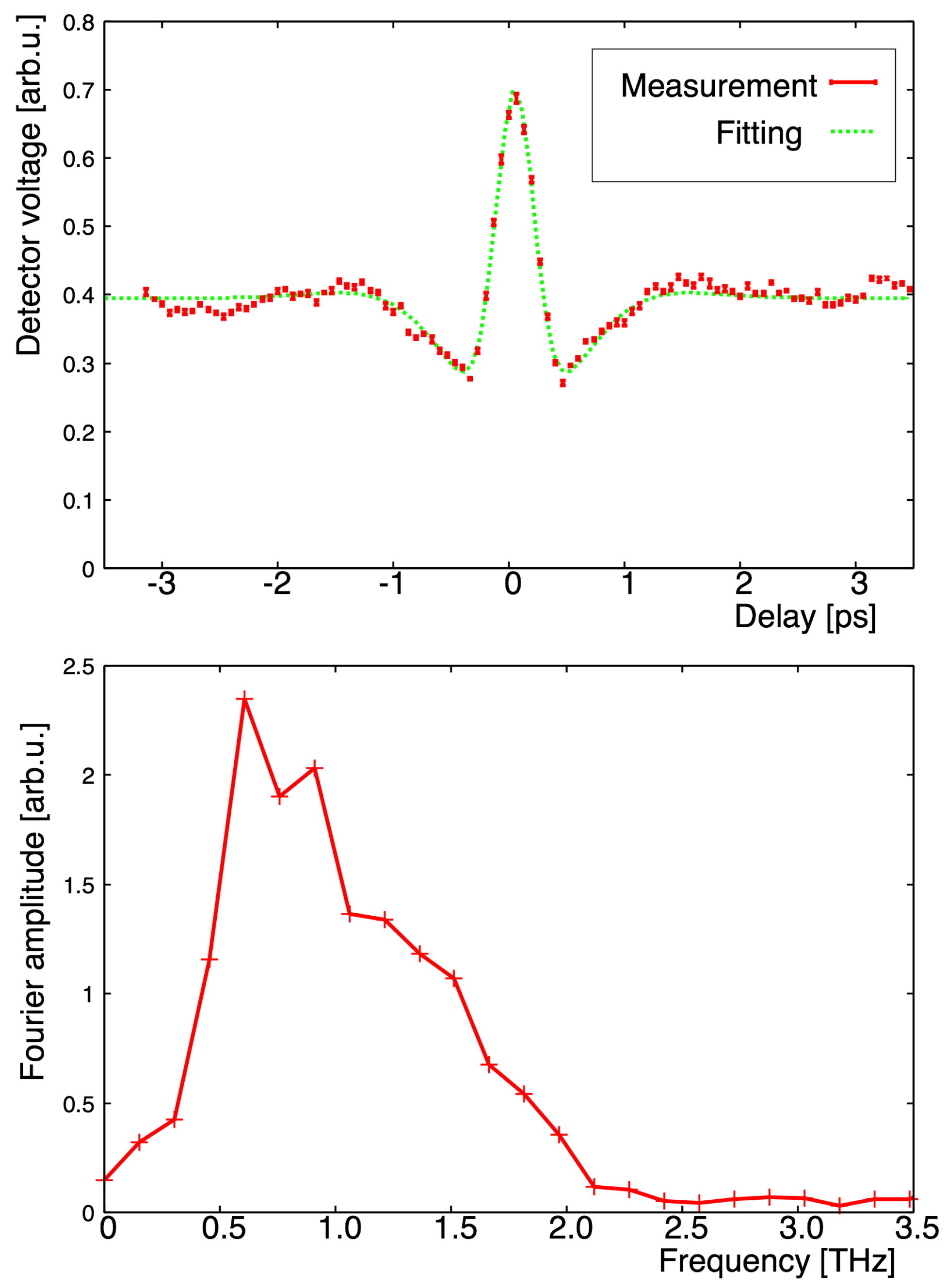}
\caption{\label{fig:bunchlengthmeasurement} Bunch length and beam spectrum
measured by the CTR interferometer system.
(Top) Interferogram of autocorrelation measurement.
(Bottom) Spectrum obtained from the interferogram.
}
\end{figure}

In order to produce coherent radiation in the THz spectrum range,
a short-bunch electron beam is necessary.
In the bunch compression mode of cERL,
a short bunch of $\sim$100 fs (rms)
can be realized at the straight section
by manipulating the longitudinal phase space 
\cite{cerlbunch}.
The original bunch length at the injector is designed to be 2 ps (rms).
The accelerating rf phase of the main linac
was shifted $+8^\circ$ from the crest
in order to imprint the time and energy correlation in the bunch.
Next, the longitudinal dispersion,
the correlation of the path length and energy
($R_{56}$), of the first arc section was adjusted
using quadrupole magnets in the arc section.
The bunch length was evaluated 
by the CTR interferometer system
at the straight section 
that was installed just behind the setup of this experiment.
The details of the CTR bunch length monitor system
and the tuning procedure for bunch compression
are given elsewhere \cite{cerlbunch}.
The measurement was performed under the same beam conditions
as this experiment,
except that
the beam focus point was shifted to the CTR target location
to minimize the effect of the beam size on the bunch length measurement.
The result is shown in Figure \ref{fig:bunchlengthmeasurement}.
The autocorrelation interferogram was fitted 
by a model function including the effects of the low-frequency cutoff.
The rms bunch length was estimated to be 120 fs.
The spectrum was obtained with a Fourier transform
and shows that the frequency reached up to 2.0 THz.
The reduction at a lower frequency range than 0.5 THz is
due to the sensitivity of the detector and the transfer line.
This confirms that THz coherent radiation can be produced in the experiment.

%
After the bunch compression tuning,
we tuned the beam optics of the straight section
to focus the beam at the cavity.
Two quadrupole magnets placed 4.78 m and 1.58 m upstream of the cavity,
Q2 and Q3 in Figure \ref{fig:straightsectionlayout}, were used.
In this beam tuning,
the cavity structure was removed from the beam axis,
and the screen monitor was inserted at the location
for beam size and position tuning.
By scanning the strength of the quadrupole magnets,
the rms beam size was minimized to  236 $\mu$m (horizontal)
and 61 $\mu$m (vertical)
at the location.

Prior to the beam experiment,
an alignment laser light was injected through the holes of the cavity mirrors
to simulate the beam path.
By inserting the screen monitor in the cavity,
the scattered laser light was recorded.
It was used as a reference
for establishing the beam path on the cavity axis.
Using the four steering magnets around the cavity,
namely, the correction coils of Q2$\sim$5, 
we prepared a combined knob to finely adjust the local beam path at the cavity
without affecting the orbit outside the area.
By scanning the local offset and angle adjustment,
we established the best beam conditions for minimizing beam loss.
Referring to the beam loss monitors,
which detect radiation produced by incidence on the mirror,
we confirmed the beam had a $\pm1$ mm clearance.
The fraction of beam loss at the cavity was estimated to be 2,500 ppm  \cite{rcdrprl}.


\section{Experimental Result}

\subsection{Observation of resonance signal}

\begin{figure}[h]
\includegraphics[width=0.9\linewidth]{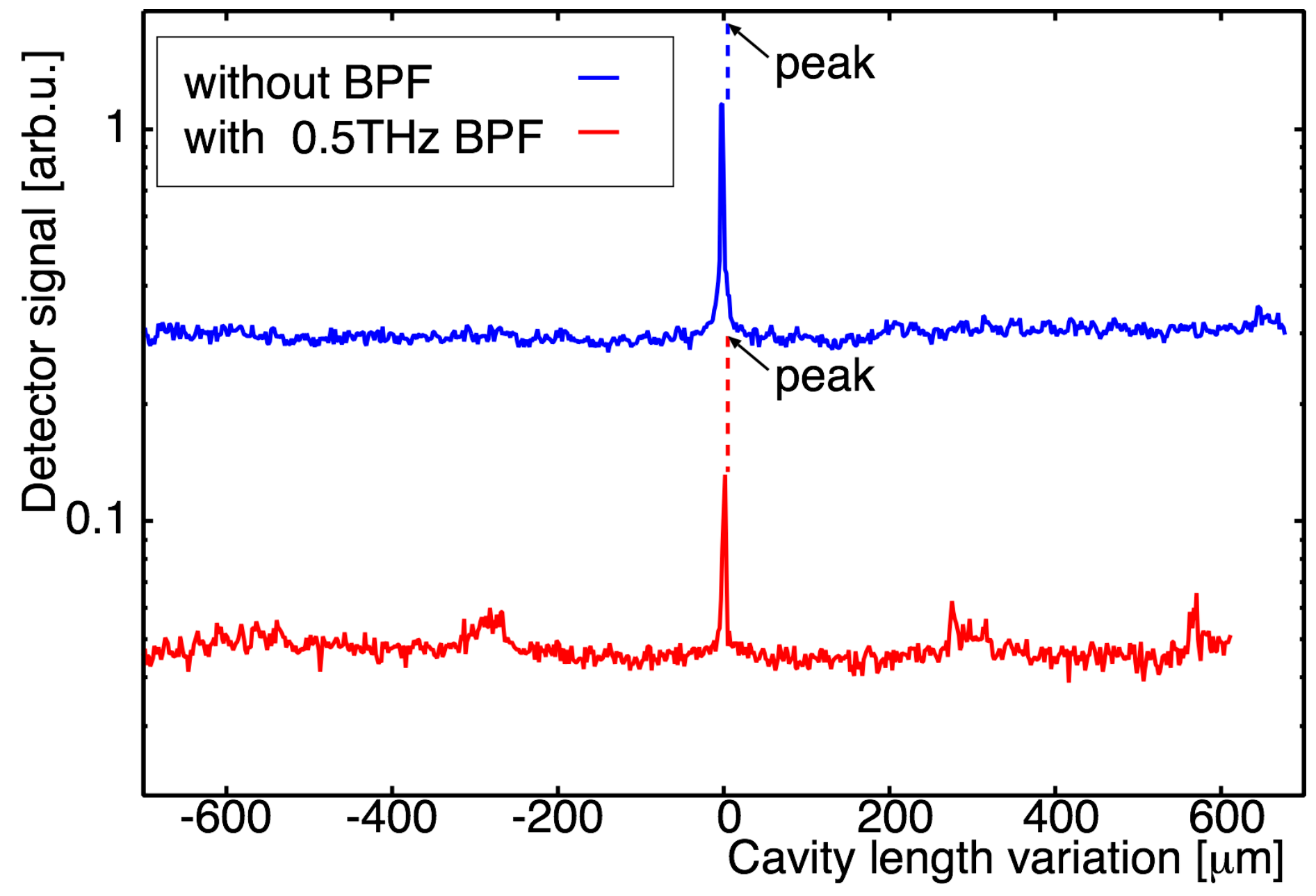}
\caption{\label{fig:resonancedata} Observation of resonance peak.
By measuring the bolometer signal while scanning the cavity length,
a strong peak appeared, showing broadband excitation.
Note that because the peak width was smaller than the sampling step
of this wide range scan,
the peak point was obtained by another fine scan data,
and it is shown by an arrow.
}
\end{figure}

The cavity length was scanned while measuring the bolometer signal.
Figure \ref{fig:resonancedata} shows the results.
The scan was performed in 2 $\mu$m steps in a 1.2 mm range.
A sharp peak was observed when the cavity satisfied
the condition of broad-band excitation.
In the case of the narrow-band measurement,
i.e., when inserting a BPF in front of the bolometer,
small bumps were repeatedly observed
with a separation corresponding to the observation frequency.
These results qualitatively agree with the discussion given in Sec.\ref{sec:broadband}.

\begin{figure}[h]
\includegraphics[width=0.9\linewidth]{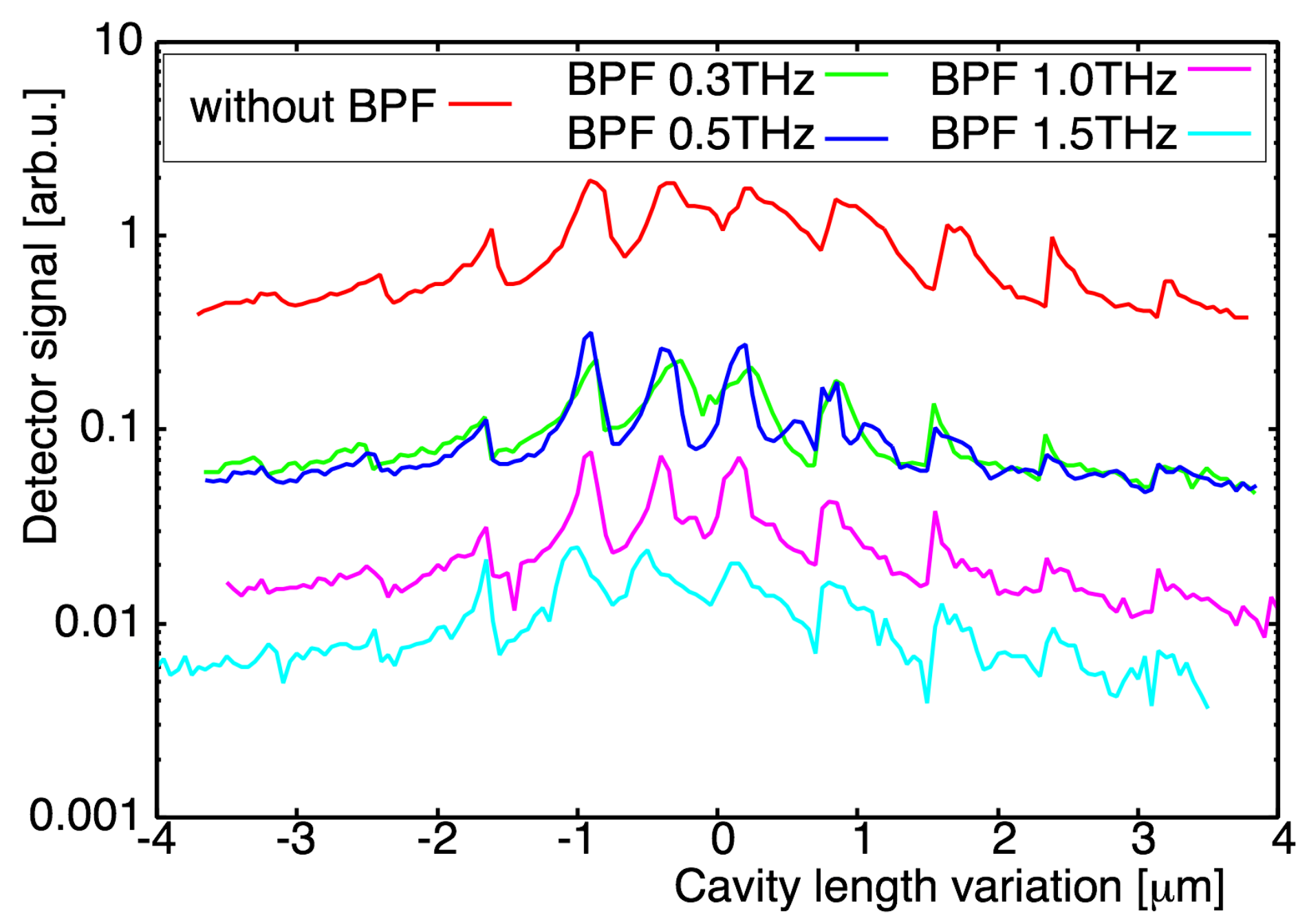}
\caption{\label{fig:scanwithfilter} Precise cavity length scan at around the resonance peak.
Results of the bolometer signal with various BPFs and without a BPF are shown.
}
\end{figure}

%
Figure \ref{fig:scanwithfilter} shows the results of 
precise scans in 50 nm steps around the main resonance peak
measured with various BPFs.
It turned out that the peak had fine structures.
The signal became weaker as the frequency increased.
It is mainly because of the original beam spectrum determined by the bunch length shown in Figure \ref{fig:bunchlengthmeasurement}.
The overall shape and positions of the fine structures were the same
in the measured frequency range of 0.3 GHz to 1.5 THz.

\begin{figure}[h]
\includegraphics[width=0.9\linewidth]{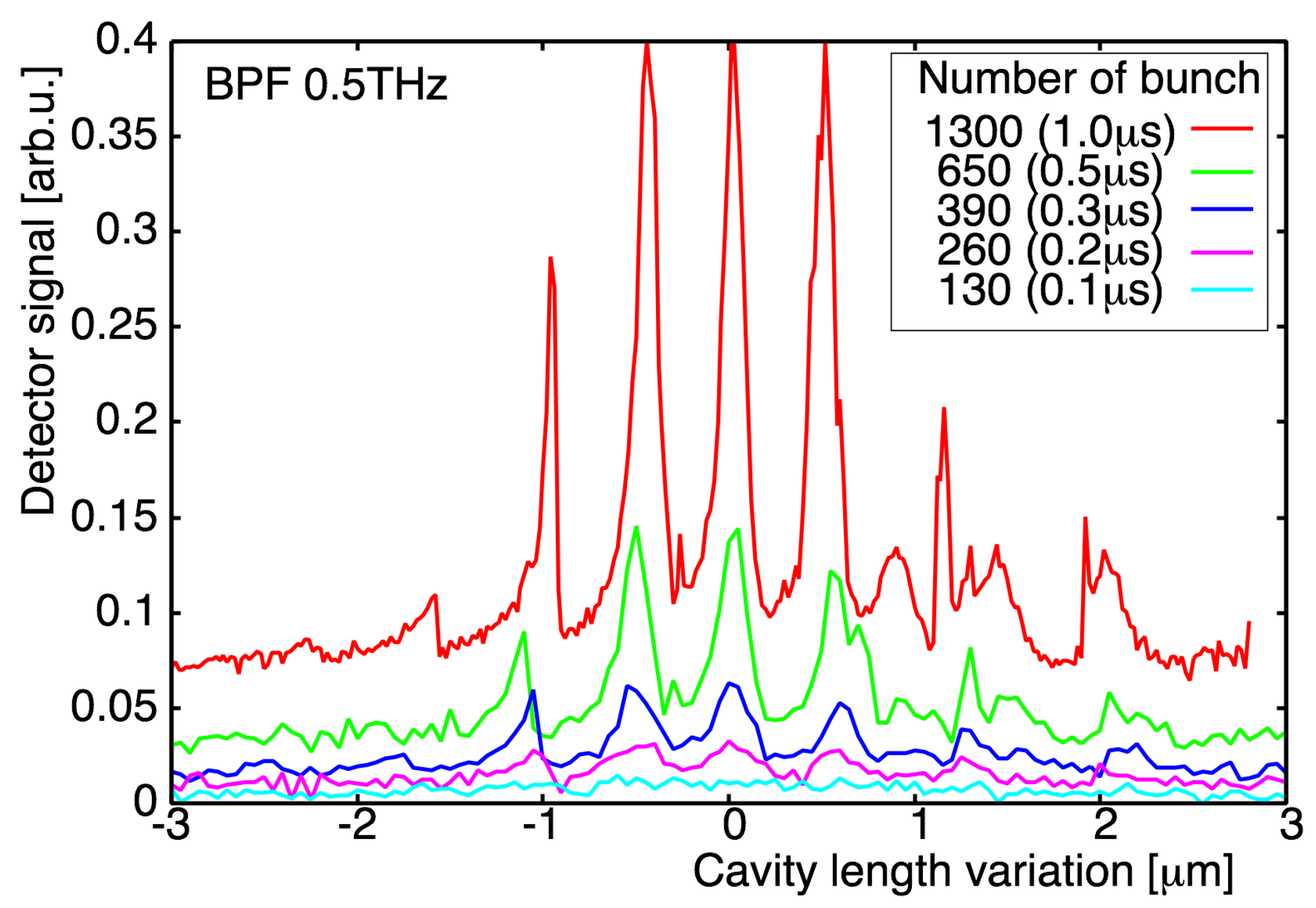}
\caption{\label{fig:bunchdependence} Precise cavity length scan obtained by
changing the number of bunches in a macro-pulse.
Measured by the bolometer with 0.5 THz BPF.
}
\end{figure}

%
In order to discuss the signal growth in a macro-pulse,
we obtained data by varying the macro-pulse length.
Figure \ref{fig:bunchdependence} shows
the results of a precise cavity length scan with 0.5 THz BPF.
As the number of bunches increases,
the intensity becomes stronger,
and the peaks in the fine structure becomes narrower.
The width of a single peak was measured  $\sim$150 nm
for the 1 $\mu$s pulse width (1300 bunch) case.

\subsection{Observation of time structure}

\begin{figure}[h]
\includegraphics[width=0.8\linewidth]{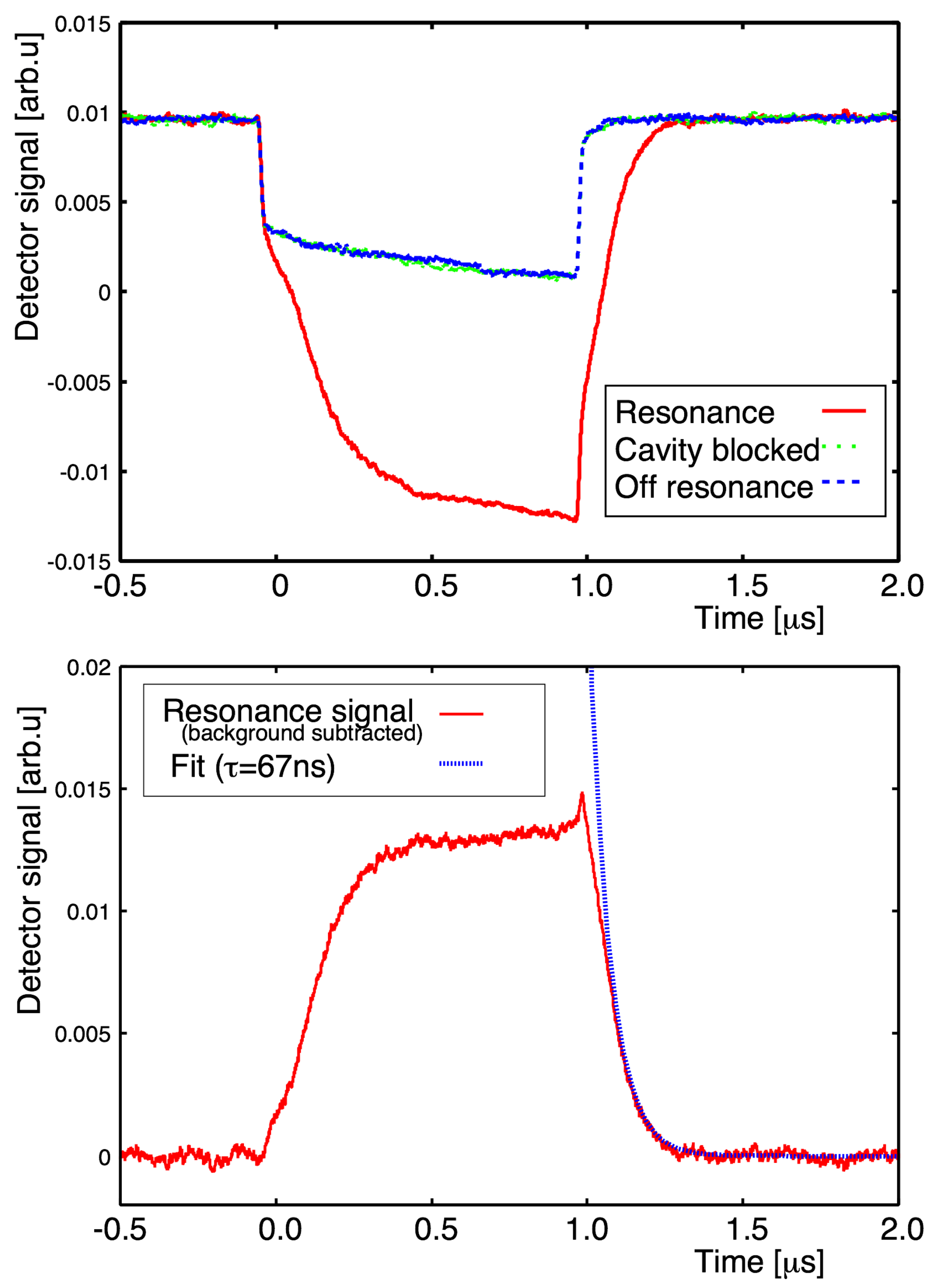}
\caption{\label{fig:waveform} Time development of THz signal measured by the QOD.
The two cases, off-resonance and blocked, 
show the contribution of the background signal.
By fitting the waveform with a decaying function,
the time constant was obtained.
}
\end{figure}

In order to directly observe the signal growth in a macro-pulse,
we recorded the waveform of the QOD signal.
By measuring the bolometer signal at the same time,
the cavity length was set at one of the resonance peaks.
As shown in Figure \ref{fig:waveform} (top),
the signal increased gradually at the beginning of the macro-pulse,
reached saturation,
and then decayed after the pulse ended.
The behavior with a time constant is an indication of resonance characteristics.
To confirm that this was a resonance signal,
following two measurements were performed simultaneously.
One was a measurement off the resonance peak, 
setting the cavity length far away from the peak.
The other was a measurement at the resonance peak
but blocking the cavity by inserting the screen monitor in the cavity.
In both of these cases,
a small signal with sharp rising and falling edges was detected.
Because these were not related to the cavity resonance,
the origin of these signals was single pass diffraction radiation
at the outer surface of the upstream cavity mirror.

Figure \ref{fig:waveform} (bottom)
shows the resonance signal obtained by 
subtracting the non-resonant background contribution.
By fitting the decaying part of the signal
with an exponential curve, $A \exp(-t/\tau)$ where $A$ and $\tau$ are free parameters,
the time constant $\tau$ was estimated to be 67 ns.
Using the relation $\tau = \frac{2 L}{c \eta}$,
we see that it corresponds to a cavity round-trip loss of $\eta=0.01$.

\subsection{Beam deceleration}

\begin{figure}[h]
\includegraphics[width=0.85\linewidth]{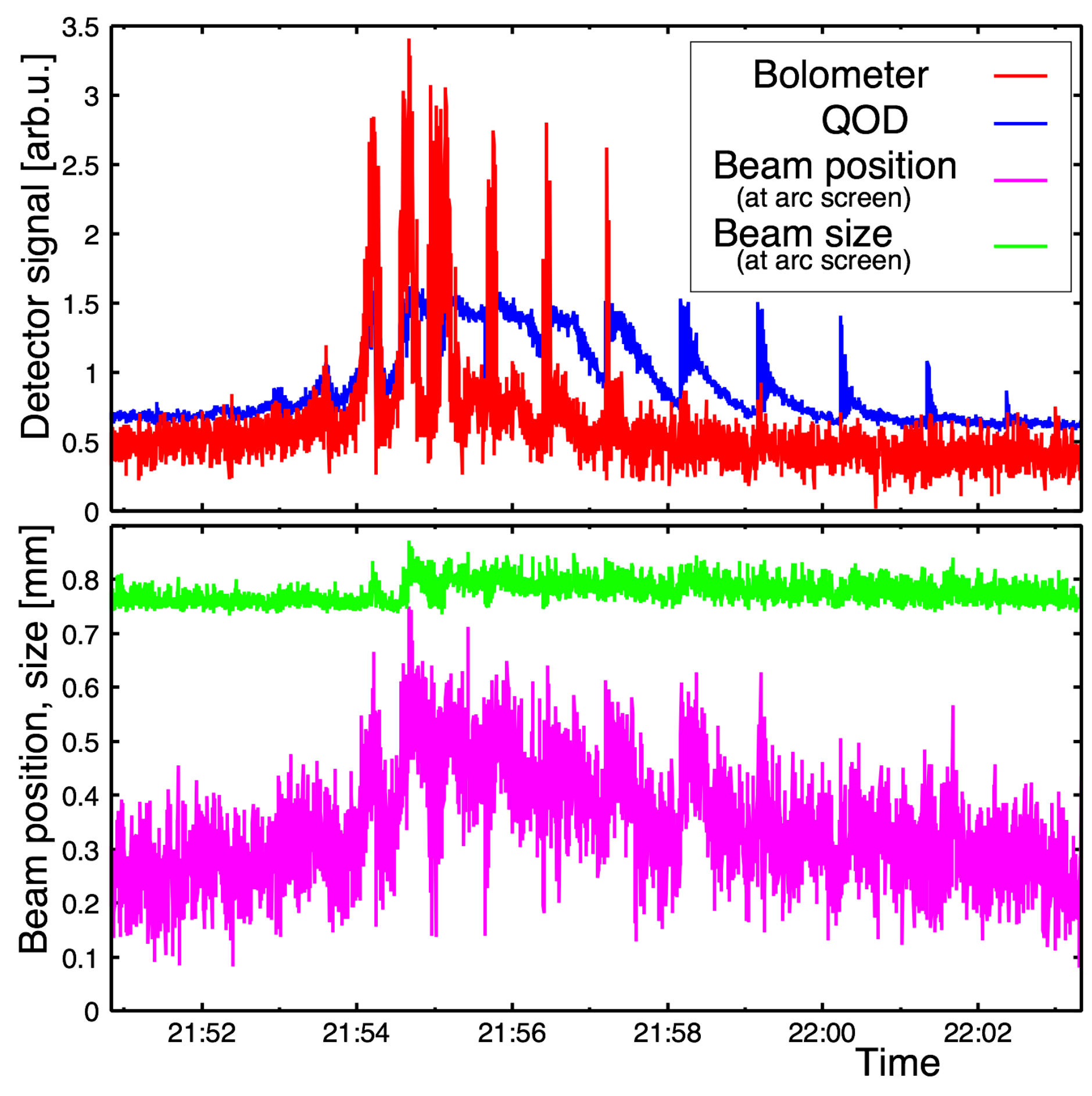}
\caption{\label{fig:dec_screen} Beam deceleration measurement by a screen monitor.
The cavity length was slowly scanned around the resonance peak.
(Top) The THz signal measured by the bolometer and the QOD.
(Bottom) Beam position and size obtained by fitting the screen monitor data.
}
\end{figure}

For the performance of a THz source,
the power of the THz signal is important.
However, a direct measurement of the absolute power
was technically difficult
because of problems related to the aperture and calibration
of the detection system.
Instead, in order to evaluate the conversion efficiency 
from the beam energy to the radiation power,
we measured the beam energy downstream of the cavity.
A screen monitor in the second arc section,
SCM2 in Figure \ref{fig:straightsectionlayout}, was used.
The horizontal dispersion at the screen monitor was designed to be 0.49 m.
We set the last quadrupole magnet in the straight section
to focus the beam horizontally at the screen to improve the energy resolution.
The horizontal rms beam size was 0.8 mm on the screen.
By fitting the recorded beam profile with a Gaussian curve
with the height, width, and the center position as free parameters,
the beam position was obtained.
Figure \ref{fig:dec_screen} shows the result
obtained while slowly scanning the cavity length.
The beam position on the screen is plotted 
with the THz signal.
It can be clearly seen that the beam position shifts to the low-energy direction
correlating with the THz radiation emission.
The maximum position shift at the resonance peak
is evaluated as 0.3 mm,
which corresponds to $6 \times 10^{-4}$ of the beam energy deceleration.

\begin{figure}[h]
\includegraphics[width=0.85\linewidth]{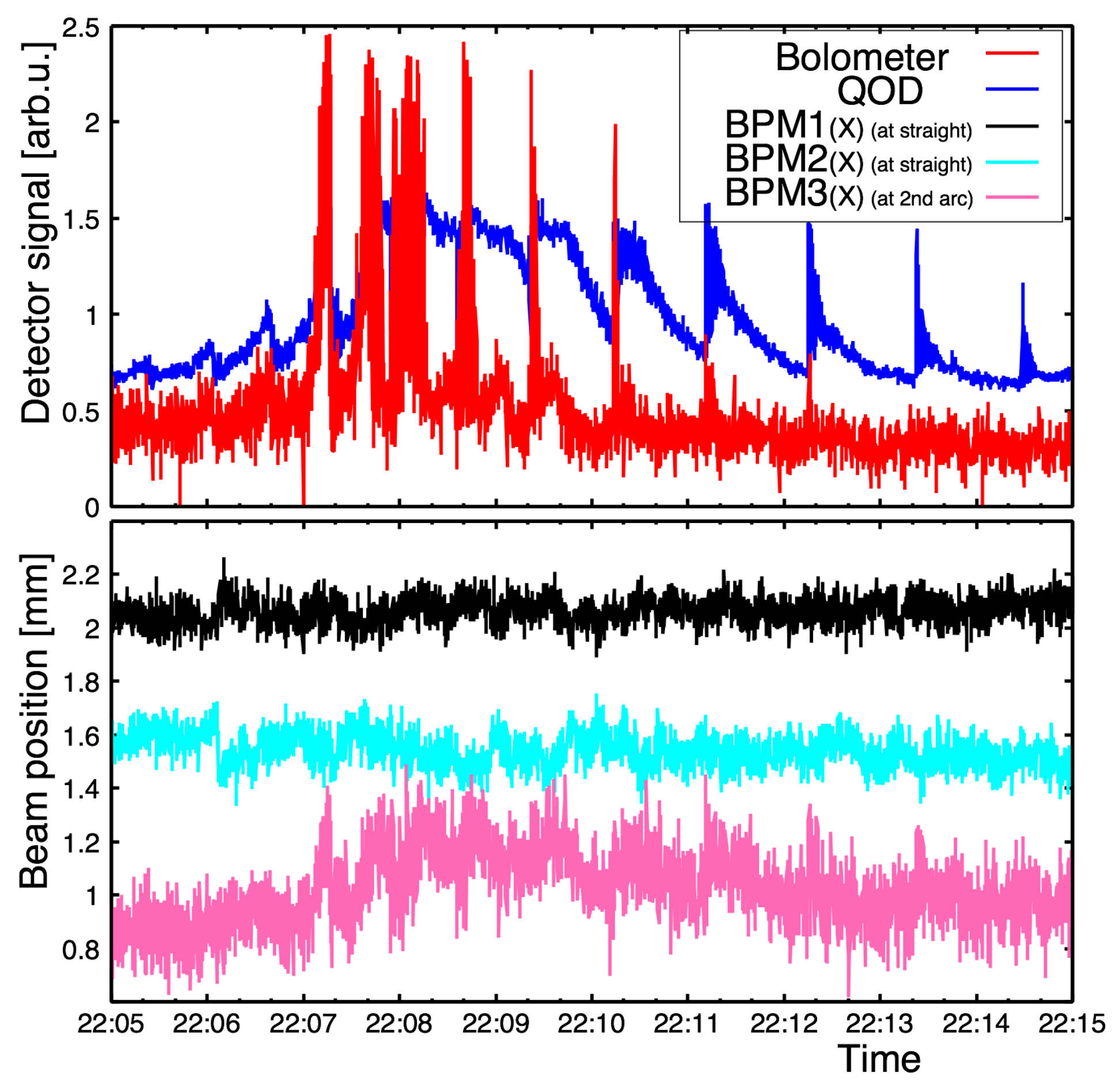}
\caption{\label{fig:dec_bpm} Beam deceleration measurement by BPMs.
The cavity length was slowly scanned around the resonance peak.
(Top) The THz signal measured by the bolometer and the QOD.
(Bottom) Beam position measured by BPMs.
BPM3 is the one in the arc section.
BPM1 and BPM2 are in the straight section.
}
\end{figure}

We repeated the same measurement using several strip-line beam position monitors (BPMs).
The BPMs are located throughout the straight section to the arc section.
As seen in Figure \ref{fig:dec_bpm},
only the BPM at the arc section (BPM3) shows a position shift,
while the BPMs at the straight section show no change.
This confirms that the beam position shift was caused by energy variation
and not by the transverse kick.


\section{Discussion}

The appearance of the resonance structure in the cavity length scan
(Figure \ref{fig:resonancedata})
proves that stimulated emission occurs in the cavity.
Strong resonance occured at the perfect synchronization between the
radiation round-trip rate and the bunch repetition rate.
This is the first result to
clearly show stimulated radiation in the THz range
in the CDR layout.

From the data obtained by changing the observation frequency range
(Figure \ref{fig:scanwithfilter}),
it was proved that 
the resonance condition coincides in the broad range of the longitudinal modes.
This agrees with the expectation from the zero-CEP shift design of the optical cavity.

The fine structures of the peaks are not what we originally expected.
We guess that each peak will correspond to higher-order transverse modes in the cavity.
Resonance conditions of higher-order transverse modes 
are degenerate in the ideal case of the confocal cavity.
However, a small perturbation,
for example, misalignment of mirrors and the effect of the mirror holes,
might split the resonance conditions.
Further studies will be needed to establish the reason conclusively.

The waveform measurement of the THz signal
(Figure \ref{fig:waveform})
confirms the resonance characteristics:
the signal growed and decayed with a time constant.
From the time constant, the cavity round-trip loss was estimated to be 
$\eta=0.01$.
On the other hand,
from the designed mode size calculated by neglecting the effects of the mirror holes,
the intensity contribution to the hole area becomes 0.02 for 0.5 THz
as a single path model.
However, this seems to be an overestimation
for the round-trip loss of the cavity eigen mode.
The result $\eta=0.01$ obtained from the waveform of the QOD
is not inconsistent with the order estimation.

The resonance width in Figure \ref{fig:bunchdependence} 
contains information on the
number of bunches whose signal stacked coherently,
and it should be related to the cavity round-trip loss.
The width of each fine structure peak becomes narrower 
as the number of bunches in the macro-pulse increases. 
However,
due to the interference of many fine peaks and the baseline background,
it is difficult to quantitatively determine the resonance width.
By roughly estimating the width as 150 nm,
the finesse turns out to be 3000 for 0.5 THz,
which cannot be explained by only 1300 bunches.

Beam deceleration due to the cavity resonance was experimentally confirmed.
This proves that 
the stimulated emission converts more beam power to the radiation 
than is obtained by just a simple multiplication by the number of bunches.
At resonance,
the decelerated energy was measured to be $6\times 10^{-4}$ 
of the beam energy.
Under the experimental conditions of 
1.2 pC bunch charge of a 17.9 MeV beam at 1.3 GHz repetition,
it corresponds to 18 W of power conversion.
This estimation neglected the contribution of the rising and falling parts
of the macro-pulse.
Although we did not measure the radiation power directly,
according to energy conservation,
the radiation power should be equal to
this energy loss of the beam.
If we operate the accelerator in the cw mode
and correct the emission of one of the directions of the cavity,
it becomes a broad-band THz source of 9 W power. 
We also note that 
as can be understood from Figure \ref{fig:mapping},
if the cavity length is intentionally shifted 
from the perfect synchronization condition,
the cavity can selectively resonate limited longitudinal modes.
In this case, it becomes a narrow-band THz source.

We note that the frequency ranges observed by the QOD and the bolometer are different.
Figure \ref{fig:dec_screen} shows 
the THz signal measured by the two detectors at the same time.
Although they share similarities,
the QOD shows wider peaks, and the overall position looks shifted.
We suppose this is because the QOD measured a lower frequency part than did the bolometer.

\begin{figure}[h]
\includegraphics[width=0.85\linewidth]{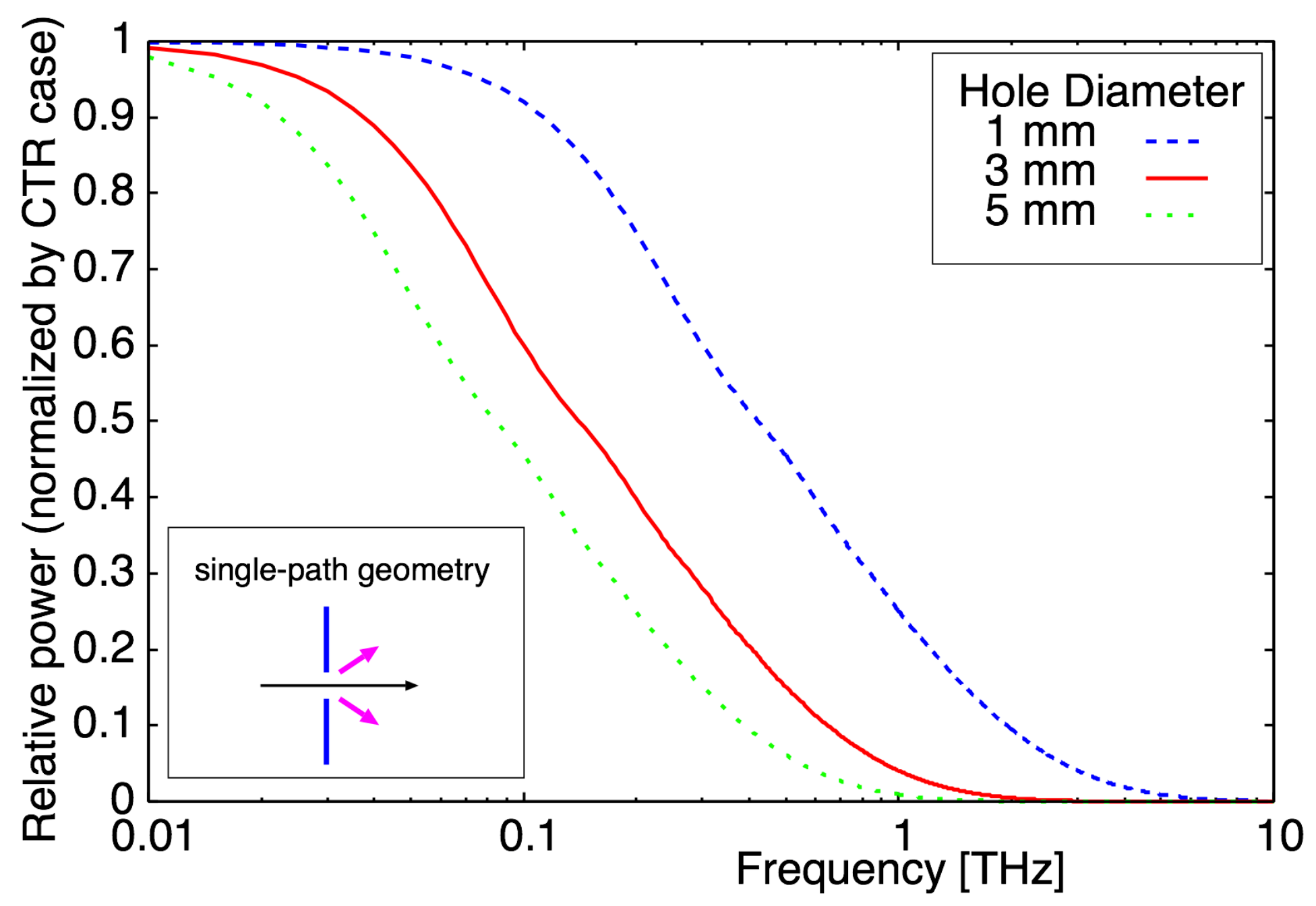}
\caption{\label{fig:drcutoff} High frequency cutoff due to the aperture.
Radiation power spectrum from a target with a circular hole
in a single pass geometry.
The aperture size gives a high-frequency cutoff.
}
\end{figure}

Here, we compare the beam deceleration result
with the calculation in Sec.\ref{sec:broadband}.
According to the beam spectrum of Figure \ref{fig:bunchlengthmeasurement},
the effective upper limit of the frequency is estimated to be 1.0 THz
although it is a smoothly decaying spectrum 
instead of the sharp cut-off of the simple model.
The round-trip loss of the cavity is chosen to be 0.01 
using the result of the waveform measurement.
The total radiation power is then calculated to be 180 W
at 1.2 pC of bunch charge.
Comparing this with the beam deceleration result,
we observe that the results differ by a factor of 1/10.
So far,
we have ignored the effects of the hole on the mirrors.
In the simple case of a single pass emission from a target,
the difference between CDR and CTR can be explained by the fact
that CDR has a high-frequency cut-off due to the aperture \cite{dr-potylitsyn}.
Figure \ref{fig:drcutoff} shows the calculation.
In the case of the 3-mm-diameter aperture,
the radiation intensity reduced by 1/10 at approximately 0.5 THz.
This effect can explain most of the difference.
Although the cut-off frequency becomes higher
as the aperture diameter becomes smaller,
there will be a trade-off against the difficulties in handling the electron beam.

%
Because the 17.9 MeV beam energy of this experiment was relatively low,
the approximation in Sec. \ref{sec:mode} 
that the beam travels at the speed of light
may need to be corrected.
The time difference for traveling the length of the cavity
with respect to the case of the speed of light
is calculated to be 150 fs,
which corresponds to 1/6.6 of the wavelength at 1 THz.
This effect will be an $\sim$20\% reduction in the excitation efficiency.

Because this was the first experiment,
we set the beam parameter at a low bunch charge of 1.2 pC,
with which it was easy to establish low-emittance and short-bunch
beam conditions.
On the other hand,
the intensity of coherent radiation will be proportional
to square of the bunch charge,
and a higher bunch charge will be more efficient 
in developing a high-power THz source.
We plan to establish beam operation at a higher bunch charge,
for example, 60 pC,
while maintaining the low-emittance and short-bunch length.
In this case,
the beam conditions may change in the macro-pulse
due to the beam loading effects in the accelerator cavities.
Further machine studies will be important.

This experiment was performed in the burst operational mode.
For a high-average-power THz source,
cw operation will be necessary.
Although we could focus the beam for 
having a $\pm$4-$\sigma$ clearance 
with respect to the aperture of the cavity,
a non-Gaussian beam halo was present that
produced beam loss at the mirrors.
In order to operate a high-current cw beam with the cavity,
management of the beam halo will be important.
Machine studies for high-current cw beam operation
have been under way.

We have demonstrated beam deceleration due to the cavity resonance.
In the energy recovery operation of an ERL,
the beam will be decelerated at the main accelerator cavities
before being dumped.
The energy variation at the dump line will be enhanced
by the gain of the main accelerator.
Sufficient energy acceptance in the dump line
will be necessary for transporting the beam to the dump without beam loss.

\section{Conclusion}

Accelerator-based THz sources
have been expected to realize high average power at broad spectrum.
A scheme based on the CDR layout is attractive,
because the layout is easy to install in a straight section,
it does not destroy the electron beam,
and it may be compatible with a high-power electron beam,
such as the one in an ERL.
Utilizing an optical cavity whose round-trip length matches
the beam bunch repetition rate,
a stimulated radiation process can be realized,
and the power extraction efficiency of the CDR can be greatly increased.

We performed an experiment with an ERL test accelerator 
that produced a low-emittance and short-bunch beam 
at a high repetition rate.
An optical cavity with a small hole was installed in the straight beam path.
A sharp resonance
in the broadband THz emission was observed in the cavity length scan.
By measuring the time structure of the THz signal,
the characteristics of resonance was confirmed.
Measurement of the beam deceleration downstream of the cavity
proved that the beam energy was converted to radiation 
efficiently by the stimulated radiation process.
This is the first experiment to clearly demonstrate
THz-range broad-band stimulated radiation in a CDR layout.

\begin{acknowledgments}

We would like to thank the cERL development team 
for their support in regard to the beam operation.
This work was partially supported by
JSPS KAKENHI Grant Number 16H05991 and 18H03473,
and by Photon and Quantum Basic Research Coordinated Development Program
from the Ministry of Education, Culture, Sports, Science and Technology, Japan.

\end{acknowledgments}


\bibliography{cdr190220submit.bbl}
\end{document}